\documentclass[aps,pra,amsmath,amssymb,onecolumn,superscriptaddress]{revtex4-2}

\usepackage{epsfig,amsmath}
\usepackage{subfigure}
\usepackage{dcolumn}
\usepackage{stmaryrd}
\usepackage{mathrsfs}
\usepackage{pifont}
\usepackage{amsthm}
\usepackage{amssymb}
\usepackage{latexsym}
\usepackage[colorlinks=true,linkcolor=blue,citecolor=blue]{hyperref}
\usepackage{color}
\usepackage{epstopdf}
\usepackage{dsfont}
\usepackage{cases}
\usepackage{multirow}
\usepackage{extarrows}
\usepackage{booktabs}
\usepackage{mathtools}

\theoremstyle{plain}

\newcommand{\la}{\langle}
\newcommand{\ra}{\rangle}

\newcommand{\ga}{\gamma}

\newcommand{\om}{\omega}
\newcommand{\Om}{\Omega}

\begin{document}
\title{Single-Photon Scattering in a Waveguide Coupled to a Giant Atom with Loss or Gain}

\author{Yu Xin}
\affiliation{College of Electronics and Information Engineering, Shandong University of Science and Technology, Qingdao 266590, Shandong, China}

\author{Jia-Ming Zhang}
\email{Email address: zhangjiaming@sdust.edu.cn}
\affiliation{College of Electronics and Information Engineering, Shandong University of Science and Technology, Qingdao 266590, Shandong, China}

\author{Bing Chen}
\email{Email address: chenbing@sdust.edu.cn}
\affiliation{College of Electronics and Information Engineering, Shandong University of Science and Technology, Qingdao 266590, Shandong, China}

\begin{abstract}
This work investigates single-photon scattering in a one-dimensional coupled-resonator waveguide interacting with a giant atom with a complex on-site energy. Within the generalized projection operator formalism, we derive analytical expressions for the scattering coefficients. We find that the interference effects of a giant atom can decouple the waveguide from the atom under specific conditions, resulting in perfect transmission, distinguishing it from a small atom. Otherwise, a giant atom with loss absorbs the incident wave, whereas a giant atom with gain not only amplifies the incident wave but also leads to scattering divergence at certain energies, corresponding to spectral singularities. We explore the critical scattering dynamics associated with these singularities, and attribute the persistent wave emission to a spectral-singularity state.  Furthermore, we show that the system with gain supports at least one time-growing bound state, which dominates the long-time dynamics. Due to the presence of these generalized bound states, the conventional time-independent scattering theory proves inadequate for such a non-Hermitian system in the gain regime. We verify our time-dependent theoretical predictions via numerical simulations of Gaussian wave packet scattering.

\end{abstract}

\maketitle

\section{Introduction}\label{intro}
Waveguide quantum electrodynamics has established a powerful framework for engineering light-matter interactions at the quantum level, offering unprecedented control over photons and quantum emitters in engineered photonic environments~\cite{blais2021,Roy2017}. With the advancement of microfabrication techniques, artificial giant atoms, such as superconducting qubits~\cite{Zheng2013,Spethmann2022}, quantum dots~\cite{Hanson2007,Xu2016}, and cold atoms~\cite{Fano1961,Szigeti2021}, can be coupled to a one-dimensional waveguide via multiple spatially separated points~\cite{Kannan2019,Vadiraj2021,WZH2020,WZH2021,Hansen2024,Sun2025b,Ma2025}. The finite propagation time of photons between coupling points and the interference effects among multiple coupling pathways lead to considerably more complex atom-waveguide interactions compared to those of small atoms. This gives rise to distinct physical phenomena, including frequency-dependent decay rates~\cite{Guo2017,Andersson2019}, Lamb shifts~\cite{Cai2021,Wang2024b}, non-Markovian dynamics~\cite{MA2024,Li2024,Sun2025}, chiral emission~\cite{Joshi2023,Roccati2024,Liu2024}, and decoherence-free interactions between atoms~\cite{Kockum2018,Carollo2020}.

For atomic systems, spontaneous emission is an inevitable process. However, its dynamics can be fundamentally controlled by modifying the local photonic density of states, or alternatively, by introducing optical gain~\cite{Sondergaard2001,Pick2017,Franke2021,Ren2024,VanDrunen2025}. Such loss or gain mechanisms are naturally described within the framework of non-Hermitian Hamiltonians via complex on-site energies. Recently, there has been growing interest in the spectral singularities of non-Hermitian systems~\cite{Spectral,Wang2016,Jin2018,gzyf-77hr}. Distinct from exceptional points, spectral singularities correspond to divergences in the continuous spectrum of scattering systems, leading to divergent transmission and reflection coefficients under incident excitation. These singularities have been extensively studied in systems with complex potentials~\cite{JHF2007,Bragg2010,Mostafazadeh2011,Abouzaid2021} and in parity-time  symmetric configurations~\cite{Heiss2013,Ramezani2014,Liu2018,Wang2024a,Wu2025}.

This paper investigates a one-dimensional coupled-resonator waveguide where a giant atom with loss or gain is coupled at two resonator sites. The single-photon scattering properties of this system are studied within a generalized projection operator formalism. We analyze the condition for the appearance of spectral singularities when introducing gain into the giant atom, and show that such singularities are associated with a spectral-singularity state embedded in the waveguide continuum. In addition to this state, we find a generalized bound state whose norm grows exponentially in time and governs the long-term dynamics.

The rest of the paper is organized as follows. In Sec.~\ref{dyna}, we introduce the model and outline the projection operator formalism. In Sec.~\ref{sec3}, we derive analytical expressions for the reflection and transmission coefficients using scattering theory and establish the condition for the emergence of spectral singularities. In Sec.~\ref{sec4}, we analyze the bound-state spectrum, numerically simulate the scattering process of a Gaussian wave packet, and interpret the resulting probability distribution in terms of the properties of the bound states. Finally, the main conclusions and outlook are presented in Sec.~\ref{con}.

\section{The Model and Projection Formalism}\label{dyna}

\begin{figure}[htbp]
\centering
\includegraphics[width=\linewidth]{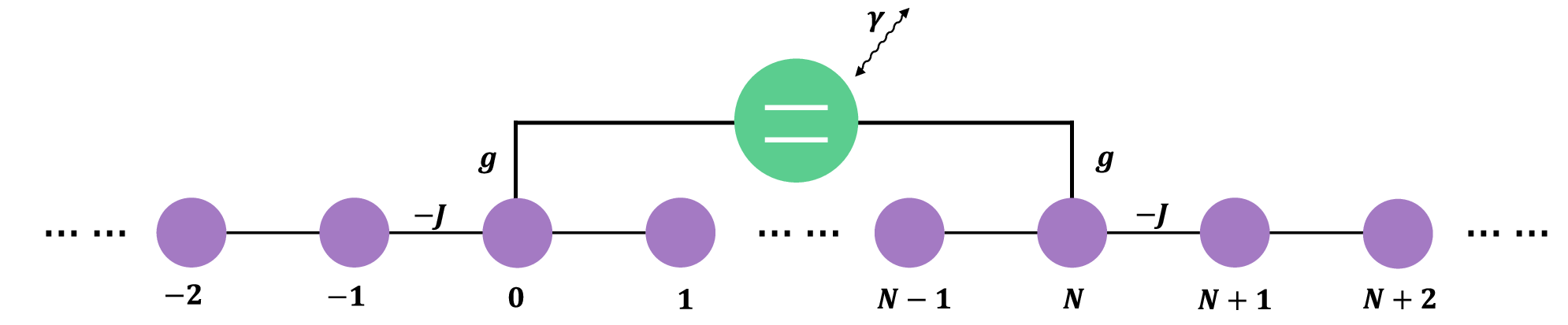}
\caption{Schematic diagram of a giant atom with loss or gain coupled to a one-dimensional coupled-resonator waveguide modeled as a tight-binding chain.}
\label{model}
\end{figure}

The system comprises an infinite one-dimensional coupled-resonator waveguide  and a two-level giant atom connected to the waveguide at two separate sites. The Hamiltonian $H$ of the total system can be divided into three parts, $H=H_a+H_c+H_I$, where
\begin{equation}\label{Ha}
H_a = (\omega_a + i\gamma)\sigma^+\sigma^-,
\end{equation}
\begin{equation}
H_c = \omega_c \sum_j c_j^\dagger c_j - J \sum_j (c_{j+1}^\dagger c_j + c_j^\dagger c_{j+1}),
\end{equation}
and
\begin{equation}
H_I = g(c_0^\dagger \sigma^- + c_N^\dagger \sigma^-) + \text{H.c.}.
\end{equation}
The Hamiltonian $H_a$ describes a two-level giant atom with transition frequency $\omega_a$, and $\sigma^\pm$ are the raising and lowering operators of the giant atom. The giant atom effectively behaves as an open cavity~\cite{Zhu2022}, with the term $i\gamma$ denoting its loss or gain, for $\gamma < 0$ in an absorptive medium~\cite{Dung2000} or $\gamma > 0$ in an amplifying medium~\cite{Raabe2008}. For the waveguide Hamiltonian $H_c$, the resonance frequency of each cavity is  $\omega_c$, the creation and annihilation operators at site $j$ are  $c_j^\dagger$  and  $c_j$, and the hopping strength between neighboring sites is  $J$. The giant atom is connected to the waveguide at two specific sites, labeled $0$ and $N$, with an equal coupling strength $g$, as shown in Fig.~\ref{model}.

Given the translational symmetry of the waveguide, we transform the Hamiltonian into the Bloch representation. The real-space operators of the waveguide can be expressed in terms of the continuous momentum basis via a Fourier transform:
\begin{equation}\label{cj}
c_j=\frac{1}{\sqrt{2\pi}}\int_{-\pi}^{\pi}c_k e^{ikj} dk,
\end{equation}
where $c_k$ is the annihilation operator for a Bloch mode with momentum $k$. Substituting Eq.~(\ref{cj}) into the waveguide Hamiltonian $H_c$ and the interaction Hamiltonian $H_I$, we have
\begin{equation}
H_c = \int_{-\pi}^\pi\omega_k c_k^\dagger c_k dk,
\end{equation}
and
\begin{equation}\label{HI}
H_I=\dfrac{g}{\sqrt{2\pi}}\int_{-\pi}^{\pi}\left[(1+e^{ikN})\sigma^+c_k+{\rm H.c.}\right]dk.
\end{equation}
The dispersion relation for the waveguide is $\omega_k=\omega_c-2J\cos k$, describing an energy band located at $\om_c$ with a width of $4J$, and the group velocity of a photon wave packet is $v_g={d\omega_k}/{dk}=2J\sin k$.

As we focus on the photon transport in the waveguide, the scattering problem is investigated using the projection operator formalism and an effective non-Hermitian Hamiltonian approach~\cite{Feshbach1962,Rotter2009,Greenberg2015,Lentrodt2020}. A brief overview of the method is provided below. It is always possible to subdivide the total space into two orthogonal subspaces via projection operators $Q$ and $P$, which satisfy the completeness relation $Q+P=1$ and the orthogonality condition $QP=PQ=0$. Accordingly, the total wave function can be expressed as $|\Psi\rangle=Q|\Psi\rangle+P|\Psi\rangle=|\Psi_Q\rangle+|\Psi_P\rangle$. Applying the projectors to the left of the stationary Schr\"{o}dinger equation $H|\Psi\rangle=E|\Psi\rangle$, we obtain two coupled equations that describe the mutual influence between the two subspace components,
\begin{gather}
  (H_{PP}-E)|\Psi_P\rangle=-H_{PQ}|\Psi_Q\rangle,\label{HE1}\\
  (H_{QQ}-E)|\Psi_Q\rangle=-H_{QP}|\Psi_P\rangle\label{HE2},
\end{gather}
where\begin{equation}\label{H}
H_{QQ}=QHQ,\ H_{PP}=PHP,\ H_{QP}=QHP,\ H_{PQ}=PHQ.
\end{equation}
After eliminating $|\Psi_P\rangle$, we obtain the Schr\"{o}dinger equation in the $Q$-subspace,
\begin{equation}\label{HE}
H_\text{eff}(E)|\Psi_Q\rangle=E|\Psi_Q\rangle,
\end{equation}
where the energy-dependent effective Hamiltonian is
\begin{equation}\label{Heff}
H_\text{eff}(E) = H_{QQ} + H_{QP} \frac{1}{E - H_{PP}} H_{PQ}.
\end{equation}
It should be emphasized that $|\Psi_Q\rangle$ represents the part of the wave function residing on the $Q$-subspace, and is not the eigenfunction with  energy $E$.

\section{Time-Independent Scattering Theory within the Projector Formalism}\label{sec3}

We assume that the $Q$-subspace is the Hilbert space of the giant atom, while the $P$-subspace spans the continuum of single-photon propagating modes in the waveguide. Since the giant atomic Hamiltonian~(\ref{Ha}) is non-Hermitian, the construction of the projector $Q$ must adhere to the general theory of non-Hermitian quantum systems. Following standard biorthogonal formulation for non-Hermitian systems~\cite{Spectral,Hatano2014,Hatano2026}, the right eigenstate $|a\rangle$ of $H_a$ and the corresponding right eigenstate $|\tilde{a}\rangle$ of its adjoint $H_a^\dag$ form a complete biorthonormal basis, satisfying $\langle \tilde{a}|a\rangle =1$. The projector $Q$ is accordingly defined as
\begin{equation}\label{Q}
Q=|a\rangle\langle \tilde{a}|.
\end{equation}
In the complementary $P$-subspace, the projection operator is expressed as
\begin{equation}\label{P}
P=\int_{-\pi}^{\pi} |k\rangle\langle k|dk,
\end{equation}
with $\langle k|k'\rangle=\delta(k-k')$.
Substituting Eqs.~(\ref{Q}) and (\ref{P}) into Eq.~(\ref{H}), we give the specific forms of the Hamiltonian components:
\begin{equation}
\begin{split}
    &H_{QQ}=(\omega_a +i\gamma)|a\rangle\langle \tilde{a}|,\\
  &H_{PP}=\int_{-\pi}^\pi\omega_k |k\rangle\langle k|dk,\\
  &H_{QP}=\dfrac{g}{\sqrt{2\pi}}\int_{-\pi}^\pi (1+e^{ikN})|a\rangle\langle k|dk,\\
  &H_{PQ}=\dfrac{g}{\sqrt{2\pi}}\int_{-\pi}^\pi(1+e^{-ikN})|k\rangle\langle \tilde{a}|dk.\label{H_PQ}
\end{split}
\end{equation}

Given that the incident wave satisfies the eigen equation $H_{PP}| k\rangle = \omega_k|k\rangle $, Eq.~(\ref{HE1}) is solved to obtain the Lippmann-Schwinger equation~\cite{ie},
\begin{equation}\label{PsiP}
|\Psi_P\rangle = |k\rangle + \frac{1}{\omega_k - H_{PP} + i\epsilon} H_{PQ}|\Psi_Q\rangle.
\end{equation}
Here, an infinitesimal term $i\epsilon$ (with $\epsilon > 0$) is introduced to shift the continuous singularities to the lower side in the complex $E$-plane, which yields a retarded Green's function describing causal propagation ($t>0$). This is equivalent to imposing the outgoing-wave boundary condition, which requires only waves propagating outward, with no incoming waves from infinity~\cite{Hatano2026}.
Substituting Eq.~(\ref{PsiP}) into Eq.~(\ref{HE2}), we have
\begin{equation}
|\Psi_Q\rangle = \frac{1}{\om_k-H_\text{eff}(\om_k)}H_{QP}|k\rangle
\end{equation}
with the effective Hamiltonian (\ref{Heff}) becoming
\begin{equation}\label{Heff+}
H_\text{eff}(\om_k) = H_{QQ} + H_{QP} \frac{1}{\om_k - H_{PP}+ i\epsilon} H_{PQ}.
\end{equation}
To simplify the analysis, we assume the resonant case $\omega_a=\omega_c$, and in what follows we take the incident wave to be right-going. The scattering-state wave function of the full system is given by
\begin{equation}\label{k}
\begin{aligned}
|\Psi(k)\rangle={}&\left[1+\dfrac{1}{\omega_k-H_\text{eff}(\omega_k)}H_{QP}+\dfrac{1}{\omega_k-H_{PP} +i\epsilon} H_{PQ}\dfrac{1}{\omega_k-H_\text{eff}(\omega_k)}H_{QP}\right]|k\rangle\\
={}&\dfrac{g(1+e^{ikN})}{\sqrt{2\pi}}\mathcal{R}(k)|a\rangle +|k\rangle+\dfrac{g^2(1+e^{ikN})}{2\pi}\mathcal{R}(k)\int_{-\pi}^\pi\dfrac{1+e^{-iqN}}{\om_k-\om_q+i\epsilon}|q\rangle dq,
\end{aligned}
\end{equation}
where $\mathcal{R}(k)$ is the inverse of $\langle \tilde{a}|\omega_k-H_{\text{eff}}(\omega_k)|a\rangle$,
\begin{equation}\label{Rk}
\mathcal{R}(k)=-\dfrac{J\sin k}{2J^2\sin k\cos k+i\gamma J\sin k-i g^2(1+e^{ikN})}.
\end{equation}
We present details of the calculation in Appendix~\ref{a:cal}. The dual state $|\tilde{\Psi}(k)\rangle$ can be obtained by replacing the Hamiltonians with their Hermitian conjugates:
\begin{equation}\label{k+}
\left|\tilde{\Psi}(k)\right\rangle
= \frac{g(1+e^{ikN})}{\sqrt{2\pi}}\tilde{\mathcal{R}}(k)|\tilde{a}\rangle
+|k\rangle + \frac{g^{2}(1+e^{ikN})}{2\pi}\tilde{\mathcal{R}}(k) \int_{-\pi}^\pi\frac{1+e^{-iqN}}{\omega_{k}-\omega_{q}+i\epsilon}|q\rangle dq,
\end{equation}
with
\begin{equation}\label{tRk}
\tilde{\mathcal{R}}(k) =-\dfrac{J\sin k}{2J^2\sin k\cos k-i\gamma J\sin k-i g^2(1+e^{ikN})}.
\end{equation}
The expressions for $|\tilde{\Psi}(k)\rangle$ and $\tilde{\mathcal{R}}(k)$ are formally identical to those for $|\Psi(k)\ra$ and $\mathcal R(k)$, respectively, apart from the substitutions $|a\ra\to |\tilde{a}\ra$ and $\ga\to-\ga$. All other factors, including the interference factors and the $i\epsilon$ prescription, remain unchanged.
In scattering theory, the $S$-matrix fully characterizes the scattering observables and connects the amplitudes of the incoming and outgoing channels. Based on the exact expression for the scattering state in Eq.~(\ref{k}), the $S$-matrix element can be expressed as~\cite{S}
\begin{equation}\label{S}
\begin{aligned}
S_{k'k}={}&\delta(k'-k) - 2\pi i \delta(\omega_{k'}-\omega_k)\langle k'|H_{PQ}|a\rangle \mathcal{R}(k) \langle \tilde{a}|H_{QP}|k\rangle\\
={}&\delta(k'-k) -\delta(\omega_{k'}-\omega_k)\dfrac{g^2 J\sin k(1+e^{-ik'N})(1+e^{ikN})}{2iJ^2\sin k\cos k-\gamma J\sin k+g^2(1+e^{ikN})}.
\end{aligned}
\end{equation}
The reflection and transmission amplitudes, $r$ and $t$, correspond to the amplitudes for the photon's momentum to be reversed or preserved upon interaction with the giant atom, respectively. These coefficients are extracted from the $S$-matrix by evaluating $S_{k'k}$ at $k' = -k$ and $k' = k$, yielding
\begin{align}
r={}&-\dfrac{g^2(1+e^{ikN})^2}{2g^2+2g^2e^{ikN}-2J\gamma\sin k+4iJ^2\sin k\cos k},\label{r}\\
t={}&1-\dfrac{g^2(1+e^{ikN})(1+e^{-ikN})}{2g^2+2g^2e^{ikN}-2J\gamma\sin k+4iJ^2\sin k\cos k},\label{t}
\end{align}
respectively. The same results can also be obtained by the stationary state approach, with the detailed calculation provided in Appendix~\ref{a:SS}.

\begin{figure}[htbp]
\centering
\includegraphics[width=\linewidth]{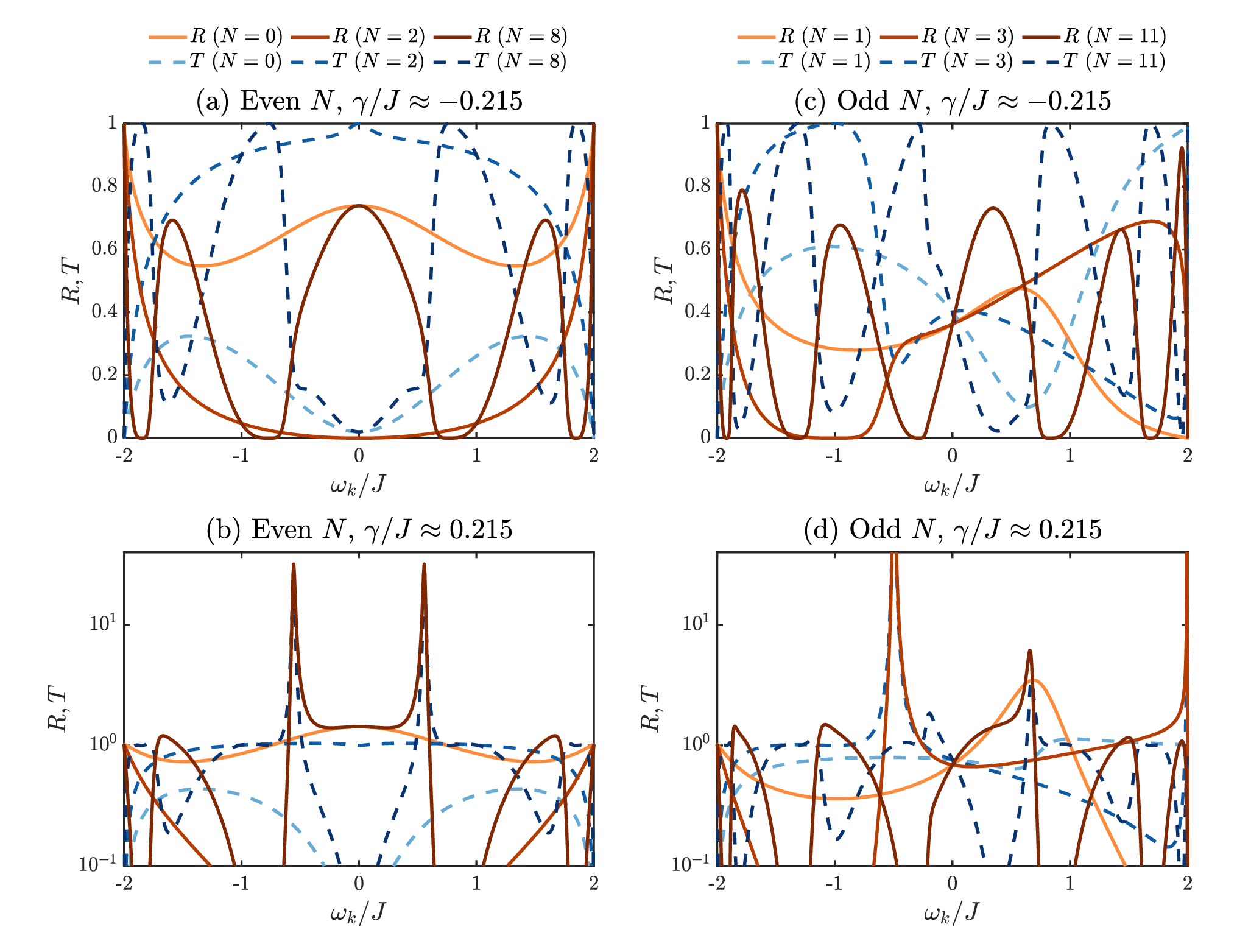}
\caption{Reflection coefficient $R$ and transmission coefficient $T$ as functions of the photon energy $\omega_k$ for different $N$. The parameters are set to $\omega_a=\omega_c=0$ and $g/J\approx0.812$. Panels (a) and (c) correspond to a loss rate $\gamma/J\approx-0.215$, while panels (b) and (d) correspond to a gain rate $\gamma/J\approx0.215$, with logarithmic vertical axes.}
\label{pic2}
\end{figure}

Figure~\ref{pic2} plots the reflection coefficient $R=|r|^2$ and the transmission coefficient $T=|t|^2$ as functions of the photon energy $\om_k$ in the waveguide for giant atoms with loss or gain.
The small-atom case corresponds to the $N=0$ case, under the substitution of the coupling strength $g \to g/2$~\cite{Tian2025}. For a small atom with loss, the sum is less than unity, whereas for gain, it is greater than unity. In contrast, the destructive interference between the two coupling points of the giant atom can result in a perfect transmission $R=0$ and $T=1$, which can be verified from Eqs.~(\ref{r}) and (\ref{t}). This occurs when, from the interaction Hamiltonian (\ref{HI}), the waveguide decouples effectively from the giant atom for wave numbers satisfying $k = (2m+1)\pi/N$ with integer $m$, irrespective of whether the giant atom with loss or gain. Except for these particular instances, which satisfy the energy-flux conservation, $R+T<1$ for $\gamma<0$, whereas $R+T>1$ for $\gamma>0$.

Furthermore, the geometric structure of the model determines the symmetry of the scattering observables, as shown in Fig.~\ref{pic2}. For even $N$ in Fig.~\ref{pic2} (a) and (b), the reflection and transmission spectra are symmetric with respect to the band center $\omega_c=0$. This symmetry originates from the interference factor $1+e^{\pm i k N}$ in the interaction Hamiltonian (\ref{HI}).
Under the transformation $k \to \pi - k$, $1 + e^{\pm i(\pi-k)N} = 1 + (-1)^N e^{\mp ikN}$.
Thus, for even $N$, the interference factor is invariant up to complex conjugation, so its modulus is unchanged. For odd $N$, it becomes $1 - e^{\mp ikN}$, whose modulus differs. From the waveguide dispersion, the symmetry about $\pi$ in momentum space maps onto the symmetry about the band center in energy space.
Consequently, for even $N$, the quantum interference between two distinct pathways, namely, direct photon propagation through the waveguide and the process of atomic absorption followed by re-emission, exhibits symmetry about $\omega_c=0$. For odd $N$, however, this symmetry is broken, giving rise to the asymmetric resonance-line shapes exemplified in Fig.~\ref{pic2} (c) and (d).

Figure~\ref{pic2} (d) reveals that the reflection probability $R$ and the transmission probability $T$ both exhibit divergence at the energy $\omega_k/J \approx-0.496$ (corresponding to $k= 1.32$) for $N=3$, a feature that has also been reported in small atom systems~\cite{Spectral,Wang2016}. Mathematically, this divergence arises from a singularity in Eqs.~(\ref{r}) and (\ref{t}), which emerges under the following conditions:
\begin{equation}\label{sp}
\left\{
\begin{gathered}
g^2 \sin {Nk} + 2 J^2 \sin k \cos k = 0,\\
g^2 + g^2 \cos {Nk} - J \gamma \sin k = 0.
\end{gathered}
\right.
\end{equation}
Notably, this condition can be satisfied only for $\gamma>0$ when $0\le k\le \pi$.

Equations~(\ref{sp}) are also the spectral-singularity conditions of the non-Hermitian model and coincide with the poles of the $S$-matrix. The divergence of $R$ and $T$ implies that bidirectional outgoing waves are sustained by the giant atom with gain, thereby invalidating standard time-independent scattering theory. To analyze scattering at these singularities, we therefore turn to the bound states and employ a time-dependent method to describe the wave-packet dynamics.

\section{Bound States and Wave-Packet Dynamics}\label{sec4}

Equation~(\ref{HE}) determines the bound state projected onto the $Q$-subspace of the giant atom~\cite{WZH2020}. By extending Bloch band theory to the generalized Brillouin zone~\cite{64,92}, the eigenvalue takes the form $E_n=\omega_c-2J\cos k_n$, where the complex wave number $k_n$ with ${\rm Im}\,k_n>0$ satisfies
\begin{equation}\label{eigenvalue}
E_n=\langle \tilde{a} |H_\text{eff}(E_n)| a \rangle
=\omega_a+i\gamma-\frac{i g^2 (1+e^{i k_n N })}{J \sin k_n};
\end{equation}
see Appendix~\ref{a:cal} for more details.

For a bound state,
\begin{equation}
|\Psi_P\rangle = \frac{1}{E_n - H_{PP} } H_{PQ}|\Psi_Q\rangle,
\end{equation}
which differs from the Lippmann-Schwinger equation, Eq.~(\ref{PsiP}).
We assume that the spectrum is non-degenerate, and the bound state for the total system  is therefore given by
\begin{equation}\label{eqmuf}
|\Psi_n\rangle = \mathcal{N}_n\left[|a\rangle+\dfrac{g}{\sqrt{2\pi}} \int_{-\pi}^\pi\dfrac{1+e^{-ikN}}{E_n-\omega_k}|k\rangle dk\right].
\end{equation}
Its dual state is obtained by replacing the energy with its complex conjugate:
\begin{equation}\label{eqmug}
|\tilde{\Psi}_n\rangle = \mathcal{N}_n^*\left[|\tilde{a}\rangle+\dfrac{g}{\sqrt{2\pi}}\int_{-\pi}^\pi\dfrac{1+e^{-ik N}}{E_n^*-\omega_k}|k\rangle dk\right].
\end{equation}
The photon wavefunction in the site representation is obtained by projecting the state vector onto the spatial basis state $\la j|$ from the left,
\begin{equation}\label{wave}
\langle j|\Psi_n\rangle=\dfrac{\mathcal{N}_n g}{2\pi} \int_{-\pi}^\pi\dfrac{e^{i k j}+e^{ik(j-N)}}{E_n-\omega_k}dk
=-\dfrac{i\mathcal{N}_n g(e^{i k_n |j|}+e^{i k_n |j-N|})}{2J\sin k_n},
\end{equation}
where we have used $\langle j|k\rangle=e^{i k j}/\sqrt{2\pi}$ and $\langle j|a\rangle=0$, and the normalization factor is given by \linebreak
$\mathcal{N}_n=1/\sqrt{1+g^2/(2\pi) \int_{-\pi}^{\pi}|1+e^{i q N}|^2/(E_n-\omega_q)^2\,dq}$ (see Appendix~\ref{a:b} for details).

In particular, for energies within the continuum $[\om_c-2J, \om_c+2J]$, one can analytically continue the energy as $E\to E+i\epsilon$ (with $\epsilon > 0$) to shift the pole off the real axis into the complex $E$-plane. For the right-going wave considered in our work ($0\le k\le \pi$), this corresponds to $k\to k+i\epsilon$. We then obtain the same form as Eq.~(\ref{eigenvalue}) in the sense of analytic continuation, which corresponds precisely to the poles of the $S$-matrix in Eq.~(\ref{S}). This indicates that the mathematical conditions for scattering divergence are intrinsically associated with the formation of the corresponding states. The unnormalized spectral-singularity state has the same form as Eq.~(\ref{eqmuf}) under analytic continuation,
\begin{equation}\label{ss}
\begin{aligned}
|\Psi_n\rangle
\sim {}&|a\rangle+\dfrac{g}{\sqrt{2\pi}}\int_{-\pi}^\pi\dfrac{1+e^{-ik N}}{E_n-\omega_k+i\epsilon}|k\rangle dk\\
={}&|a\rangle+\dfrac{g}{\sqrt{2\pi}}\mathcal P\int_{-\pi}^\pi\dfrac{1+e^{-ik N}}{E_n-\omega_k}|k\rangle dk-\dfrac{i\pi g}{\sqrt{2\pi}} \left(\frac{1+e^{-ik_n N}}{2 J \sin k_n}|k_n\ra+\frac{1+e^{ik_n N}}{2 J \sin k_n}|-k_n\ra\right),
\end{aligned}
\end{equation}
which approximates a bidirectional plane wave in the waveguide~\cite{Spectral,Mostafazadeh2009}. However, since the Green's function diverges at the spectral singularity, the spectral-singularity state requires a generalized normalization constant.

According to Eq.~(\ref{wave}), bound states with complex energies $E_n$ outside the continuum decay exponentially as $|j|\to \infty$. Their eigenenergies reside on the first (physical) Riemann sheet, which corresponds to poles in the upper half of the complex $k$-plane.
The conventional bound states exist below and above the energy band, lying on the positive imaginary $k$ axis (${\rm Re}\,k=0$) and the positive part of the vertical line (${\rm Re}\,k=\pi$), respectively. From the solvability condition of Eq.~(\ref{eigenvalue}), one can find that the existence of these conventional bound states is strictly limited to the case $\ga=0$. Because the energy of photons propagating in the waveguide falls within the continuum of the waveguide band, they do not resonate with these discrete bound states. Consequently, the scattering process is completely unaffected by these localized states.
Bound states with energies in the upper half of the complex $E$-plane are the so-called time-growing states, whose norms grow exponentially with time~\cite{gzyf-77hr,Li2019,Xiao2021,Zhang2025}. They occupy the first quadrant of the complex $k$-plane within the first Brillouin zone. Their counterparts with energies in the lower half of the complex $E$-plane are the time-decaying states, whose norms decay exponentially. These time-decaying states lie in the second quadrant of the complex $k$-plane.

The first Riemann sheet is connected to the second Riemann sheet through a branch cut along $\om_c-2 J\le E\le \om_c+2 J$, with branch points at $E=\om_c\pm 2 J$. This second sheet is referred to as unphysical because the eigenfunctions of its states diverge away from the giant atom. On the branch cut, spectral-singularity states can appear. They are also termed the delocalized bound states in the continuum~\cite{Longhi2014}, or alternatively, resonances in the continuum~\cite{64}. The conventional localized bound states in the continuum can appear in Hermitian systems and are characterized by a localized spatial profile and vanishing coupling to the continuum. In contrast, these spectral-singularity states arise from the non-Hermitian structure of the system and exhibit a wave function that is extended throughout the spatial extent of the system, forming a plane-wave-like solution given by Eq.~(\ref{wave}).

\begin{figure}[htbp]
\centering
\includegraphics[width=\linewidth]{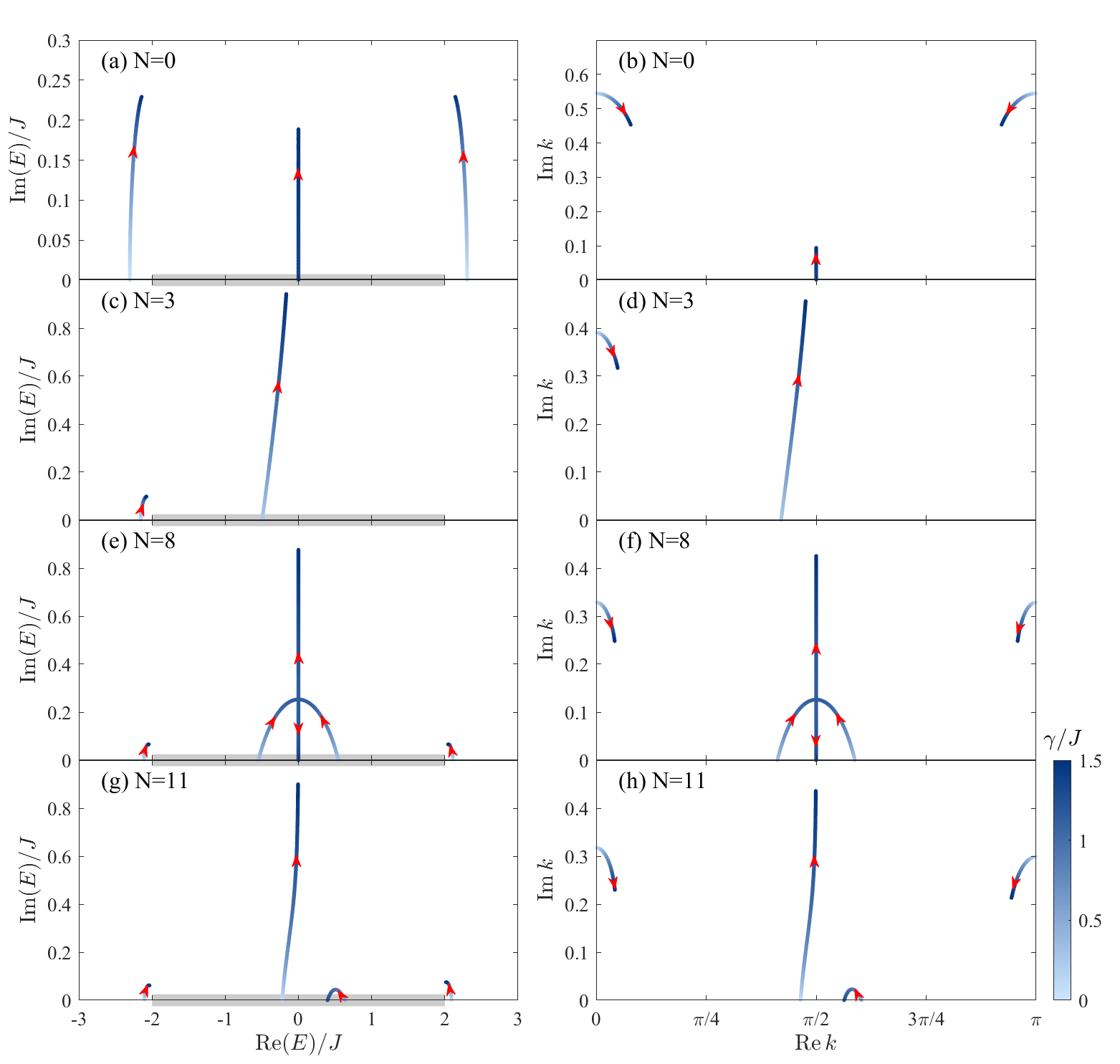}
\caption{Trajectories of the complex energies on the upper half of the first Riemann sheet (left panels) and the corresponding wave numbers in the first quadrant (right panels) for different $N$ as functions of the gain rate $\gamma$. The color bar indicates \(\gamma\), with its value increasing from light to dark. Arrows along the trajectories show the direction of increasing $\gamma$. The parameters are set to $\omega_a=\omega_c=0$ and $g/J\approx0.812$, identical to those in Fig.~\ref{pic2}.}
\label{Trajectories}
\end{figure}

Trajectories of the complex energies on the upper half of the first Riemann sheet and the corresponding wave numbers in the first quadrant for $\gamma\ge 0$ are displayed in Fig.~\ref{Trajectories}. For $\gamma<0$, the system is dissipative, so the energies lie in the lower half of the complex $E$-plane, and no wave numbers appear in the first quadrant of the complex $k$-plane. According to Eq.~(\ref{eigenvalue}), changing the sign of $\gamma$ corresponds to a reflection about the real axis in the $E$-plane and about the imaginary axis in the $k$-plane. This symmetry allows the solutions for $\gamma<0$ in the lower half of the $E$-plane to be inferred from those in the upper half, and the corresponding wave numbers in the second quadrant to be inferred from those in the first quadrant.
Figure~\ref{Trajectories} shows that increasing $\gamma$ converts the conventional bound states present at $\gamma=0$ into time-growing states. As $\gamma$ increases further, additional spectral-singularity states emerge and transition into time-growing states.
For even $N$, Fig.~\ref{Trajectories} demonstrates that the trajectories exhibit symmetry about $E=0$ in the energy plane and about $k = \pi/2$ in the momentum plane. For odd $N$, this symmetry is broken. This phenomenon is analogous to that shown in Fig.~\ref{pic2}, arising from the interference factor $1+e^{\pm i k N}$ in the interaction Hamiltonian (\ref{HI}), as we have shown in our analysis of Fig.~\ref{pic2}.

For the small atom scenario in Fig.~\ref{Trajectories} (a) and (b), both lower and upper localized bound states exist at $\ga=0$. The gain value $\gamma/J=2g^2/J^2\approx1.317$ marks a critical point beyond which a third solution emerges. At this critical gain, one eigenenergy lies inside the energy band, signaling the emergence of a spectral-singularity state.
Figures~\ref{Trajectories} (c) and (d) correspond to the $N=3$ case, where only one lower localized bound state exists at $\ga=0$. There is also a spectral-singularity state at energy $E/J\approx-0.496$ (corresponding to $k = 1.32$), whose dynamics we will analyze in the following. Although sharing the same symmetry with the small atom model, the $N=8$ case exhibits an exceptional point at $\ga/J\approx1.145$ in Figs.~\ref{Trajectories}(e) and (f), accompanied by additional solutions. 
At this exceptional point, the two complex solutions coalesce into a single solution whose eigenvalue is purely imaginary with a positive imaginary part. Consequently, the bound-state expressions in Eqs.~(\ref{eqmuf}) and (\ref{eqmug}), which are derived under the assumption of a non-degenerate spectrum, are no longer applicable. A detailed treatment of this regime is beyond the scope of our work.
For $N=11$ in Figs.~\ref{Trajectories} (g) and (h), more solutions appear, along with an upper localized bound state at $\ga=0$, compared with the likewise asymmetric $N=3$ case.

We now investigate the dynamical evolution of a wave packet in the waveguide after interacting with a giant atom with loss or gain. We assume that at time $t=0$, when the wave packet encounters the giant atom, it has the following form in momentum space,
\begin{equation}\label{initialk}
|\phi(0)\rangle = \int_{-\pi}^\pi \beta(k)|k\rangle dk.
\end{equation}
Taking into account the energy eigenstates including both the discrete bound states and continuous scattering states, this state can be expanded as:
\begin{equation}\label{phi0}
|\phi(0)\rangle=\sum_n C_n |\Psi_n\rangle+\int_{-\pi}^\pi D(k) |\Psi(k)\rangle dk,
\end{equation}
with
\begin{align}
C_n ={}& \langle \tilde{\Psi}_n | \phi(0) \rangle
=\dfrac{g\mathcal{N}_n}{\sqrt{2\pi}}\int_{-\pi}^\pi\dfrac{\beta(k)(1+e^{ikN})}{E_n-\omega_k-i\epsilon}dk, \label{Cn} \\
D(k) ={}& \langle \tilde{\Psi}(k) | \phi(0) \rangle
=\beta(k) +\dfrac{g^2(1+e^{-ikN})}{2\pi}\tilde{\mathcal{R}}^*(k)\int_{-\pi}^\pi \dfrac{\beta(q)(1+e^{iqN})}{\omega_k-\omega_q-i\epsilon}dq,\label{Dk}
\end{align}
representing the expansion coefficients at $t=0$. To avoid the need to interpret the completeness relation in a distributional sense due to the spectral-singularity state in Eq.~(\ref{ss})~\cite{Andrianov2010}, in what follows we only analyze bound states satisfying Eq.~(\ref{eigenvalue}) with ${\rm Im}\,k_n>0$. A bound state with a sufficiently small, positive imaginary part ${\rm Im}\,k_n>0$ can be used to approximate the nearby spectral-singularity state. The completeness relation is proved in Appendix~\ref{a:i}.

The temporal evolution of the state vector $|\phi(t)\rangle$ governed by the time-dependent Schr{\"o}dinger equation,
\begin{equation}\label{tdse}
i\dfrac{d}{dt}|\phi(t)\rangle=H|\phi(t)\rangle,
\end{equation}
yields
\begin{equation}
|\phi(t)\rangle = e^{-iHt} |\phi(0)\rangle=\sum_n C_n e^{-i E_n t}|\Psi_n\rangle+\int_{-\pi}^\pi D(k) e^{-i\omega_kt}|\Psi(k)\rangle dk.
\end{equation}

By projecting the photon wave packet onto the spatial basis $\langle j|$ from the left, we obtain its configuration-space representation and the probability amplitude at the $j$-th site:
\begin{equation}\label{jphi}
\langle j|\phi(t)\rangle=\sum_n C_n e^{-i E_n t}\langle j|\Psi_n\rangle+\int_{-\pi}^\pi D(k) e^{-i\omega_kt}\langle j|\Psi(k)\rangle dk.
\end{equation}
According to the stationary phase method~\cite{Carl1999}, the integral in Eq.~(\ref{jphi}) vanishes in the long-time limit, as elaborated in Appendix~\ref{a:lt}.
Consequently, only contributions from the bound states persist. The time evolution of the bound state is described by
\begin{equation}\label{psi}
\Psi_n(j,t)=C_n e^{-i E_n t}\langle j|\Psi_n\rangle
=-\frac{i\mathcal{N}_nC_n g}{2J\sin k_n}e^{-iE_n t }
(e^{i k_n |j|}+e^{i k_n |j-N|}).
\end{equation}
Equation~(\ref{psi}) corresponds to the time evolution of the eigenstate under the Siegert boundary condition~\cite{Siegert}, which we established in Appendix~\ref{a:sb}. It reveals that the imaginary part of the wave number, ${\rm Im}\,k_n$, governs the spatial behavior and the imaginary part of the energy ${\rm Im}\,E_n$ controls the temporal evolution. This is illustrated by the example that follows.

To directly visualize the time evolution of the probability distribution, we simulate the propagation of a single-photon wave packet~\cite{Spectral,Wang2016}. The wave packet is initialized at time $t=-t_c$ as a normalized Gaussian distribution of
site occupation amplitudes,
\begin{equation}
\langle j|\phi(-t_c)\rangle=\pi^{-\frac{1}{4}}\alpha^{\frac{1}{2}}{e^{-\frac{\alpha^2(j-j_c)^2}{2}}}e^{i k_c j}.
\end{equation}
Here, $j_c$ and $k_c$ represent the mean position and mean wave number of the wave packet, respectively. The wave packet is initialized sufficiently far from the coupling points ($j_c\ll 0$). Its mean group velocity is $v_c=2J\sin k_c$, and $t_c=-j_c/v_c$ corresponds to the arrival time of the peak at the giant atom. The parameter $\alpha$ controls the spatial width of the wave packet. A larger $\alpha$ corresponds to a narrower spatial profile and, through the Fourier transform relation, a broader momentum distribution, and vice versa.

\begin{figure}[htbp]
\centering
\includegraphics[width=0.9\linewidth]{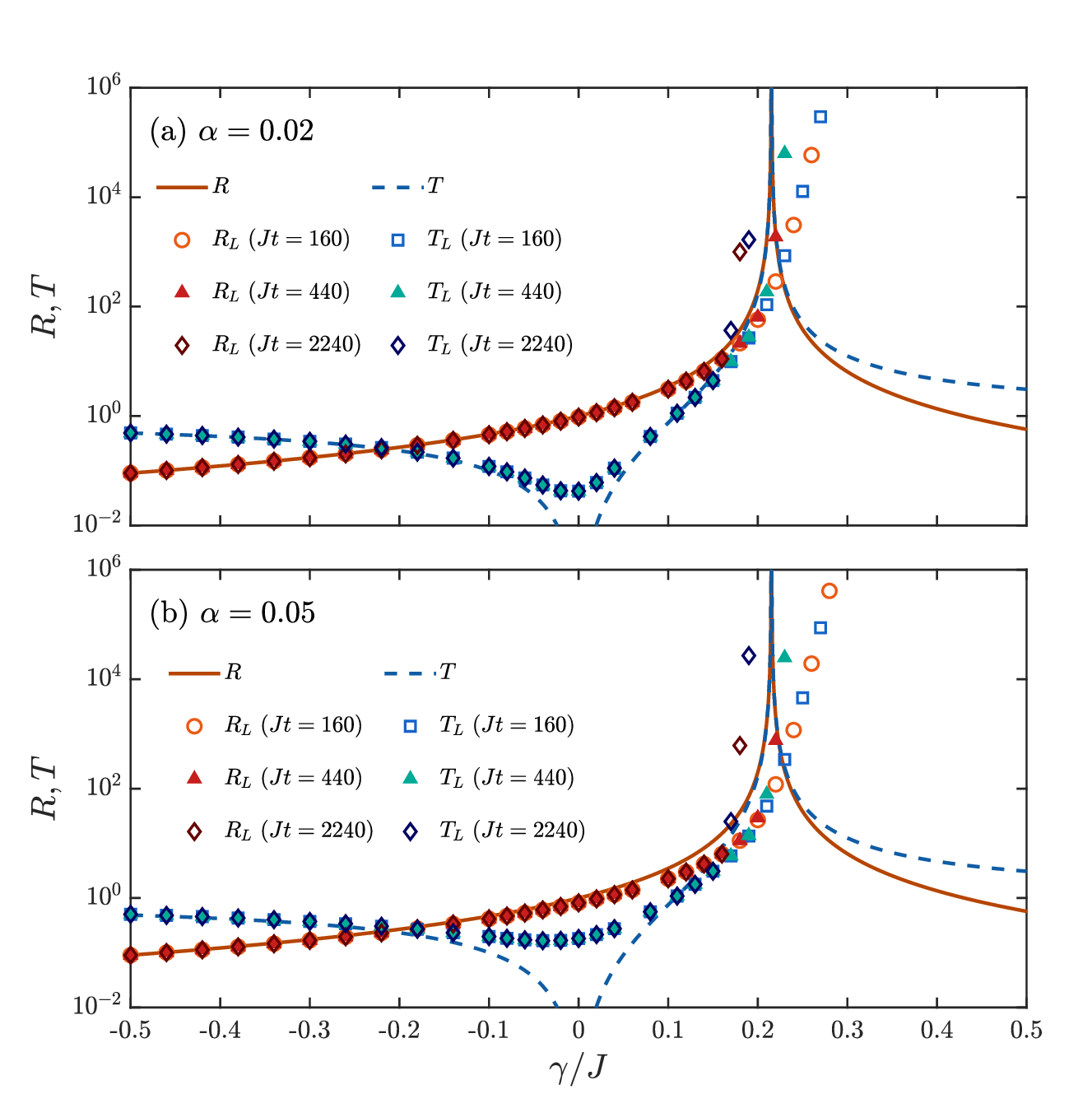}
\caption{Reflection and transmission coefficients for $N=3$ with fixed wave number $k=1.32$ as functions of $\ga$. The lines are derived from the time-independent scattering method, while the discrete points are obtained from Gaussian wave-packet dynamics at times $Jt = 160$, $440$, and $2240$.
The incident wave packet is initially localized at $j_c = -500$ on a lattice of $10^4$ sites, with (a) $\alpha = 0.02$ and (b) $\alpha = 0.05$.
Other parameters are set to $\omega_a=\omega_c=0$ and $g/J\approx0.812$, identical to those in Figs.~\ref{pic2} and \ref{Trajectories}.}\label{RT}
\end{figure}

Substituting the wave function expanded in the Wannier basis, $|\phi(t)\rangle = \phi(a,t) |a\rangle + \sum_j \phi(j,t) |j\rangle$, into the time-dependent Schr{\"o}dinger equation~(\ref{tdse}), a set of coupled equations for the coefficients can be derived as
\begin{equation}\label{num}
\begin{dcases}
i\dfrac{d\phi(j,t)}{dt} = \omega_c \phi(j,t)-J[\phi(j+1,t)+\phi(j-1,t)]+g (\delta_{j,0}+\delta_{j,N})\phi(a,t), \\
i\dfrac{d\phi(a,t)}{dt} = (\omega_a+i\gamma)\phi(a,t)+g[\phi(0,t)+\phi(N,t)].
\end{dcases}
\end{equation}
Based on convergence checks, Eq.~(\ref{num}) are integrated numerically using a fourth-order adaptive Runge-Kutta method with a relative error tolerance of $10^{-6}$ and an absolute error tolerance of $10^{-8}$. A convergence test with the tolerance relaxed by a factor of $10$ shows that the logarithmic probability amplitude has a maximum relative deviation of about $5\%$ at long times, and the long-time data are used to extract the effective growth rate from its slope. We define the reflection and transmission coefficients by summing all probability amplitudes propagating to the left and to the right, respectively, as follows~\cite{Chen2011,gzyf-77hr}:
\begin{equation}
R_L(t) = \sum_{j < 0} P(j,t), \qquad T_L(t) = \sum_{j > N} P(j,t),
\end{equation}
where $P(j,t)$ represents the occupation numbers at each lattice site $j$ as a function of time $t$.
In the simulation, we set the incident wave packet with mean wave number $k_c=1.32$, initially localized at $j_c = -500$ on a lattice of $10^4$ sites with open boundaries. Since the maximum group velocity of the wave packet in the one-dimensional tight-binding waveguide is $2J$, the maximum distance the wave packet can travel during our simulation time $Jt = 2500$ is approximately $5000$ sites. Therefore, the wave packet remains away from the lattice boundaries, naturally eliminating any boundary reflection effects. For the $N=3$ case, the resulting reflection and transmission coefficients as functions of $\ga$ are shown in Fig.~\ref{RT}.

In Fig.~\ref{RT}, the time-dependent wave-packet simulations agree well with the time-independent analytical expressions based on Eqs.~(\ref{r}) and (\ref{t}) in the loss regime. However, near the critical point $\gamma=0$, a deviation appears in the transmission coefficient. This is because the transmission coefficient for $k=1.32$ drops sharply in this region, while the simulated Gaussian wave packet has a finite width in momentum space. The deviation near the spectral singularity originates from the same momentum-space broadening of the simulated wave packet. Comparing Figs.~\ref{RT} (a) and (b), the discrepancies can be minimized by using a smaller value of $\alpha$.

In the gain regime, due to the emergence of time-growing bound states, both the reflection and transmission coefficients of the simulated wave packet increase with time, making the time-independent method inapplicable. Comparing the three time points $Jt=160$, $440$, and $2240$ in Fig.~4, one can see that for larger $\gamma$, the reflection and transmission coefficients grow faster with time, and the amplification effect becomes more pronounced. This is because a larger $\gamma$ corresponds to stronger gain. As shown in Fig.~\ref{Trajectories}, the imaginary part of the energy of the time-growing state increases with $\gamma$, which means that the norm of the state grows faster in time.

\begin{figure}[htbp]
\centering
\includegraphics[width=\linewidth]{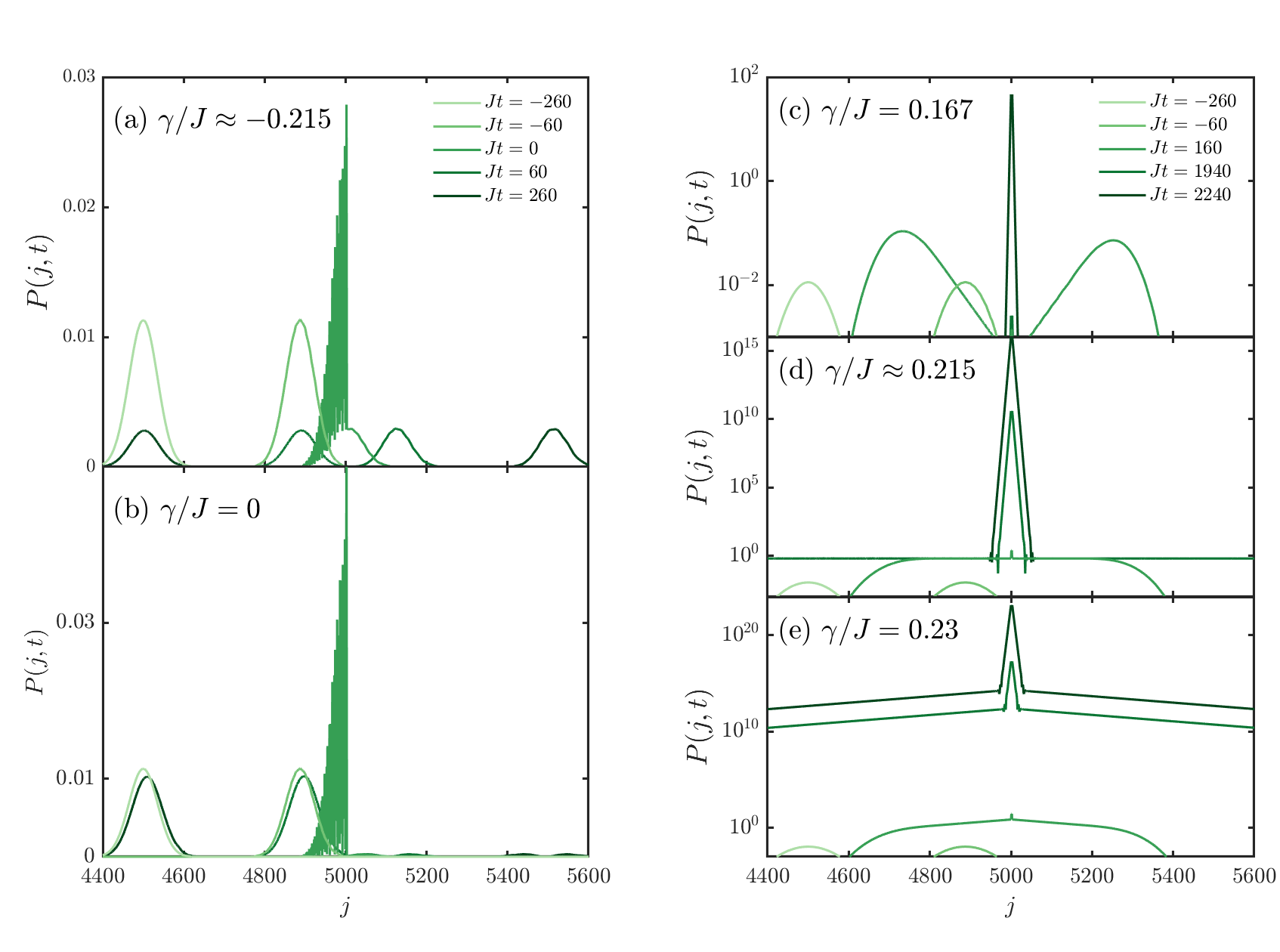}
\caption{Scattering of a Gaussian wave packet ($k_c=1.32$) for $N=3$ at (a) $\ga\approx -0.215$, (b) $\ga=0$, (c) $\ga=0.167$, (d) $\ga\approx0.215$, and (e) $\ga=0.23$. The incident wave packet is initially localized at $j_c = -500$ on a lattice of $10^4$ sites, with $\alpha = 0.02$.
Other parameters are set to $\omega_a=\omega_c=0$ and $g/J\approx0.812$, identical to
those in Figs.~\ref{pic2}-\ref{RT}.}\label{fig5}
\end{figure}

Figures \ref{fig5} (a)-(e) illustrate the evolution of a Gaussian wave packet. At the initial time $t=-t_c$, the momentum distribution of the wave packet is also Gaussian,
\begin{equation}
|\phi(-t_c)\ra=\pi^{-\frac{1}{4}}\alpha^{-1/2}\int_{-\pi}^\pi e^{-\frac{(k_c-k)^2}{2\alpha^2}+i(k_c-k)j_c}|k\rangle dk.
\end{equation}
The time evolution of the wave packet before encountering the giant atom can be expressed as
\begin{equation}\label{phit}
|\phi(t)\ra=\pi^{-\frac{1}{4}}\alpha^{-1/2}\int_{-\pi}^\pi e^{-\frac{(k_c-k)^2}{2\alpha^2}+i(k_c-k)j_c}e^{-i\omega_k (t+t_c)}|k\rangle dk.
\end{equation}
If the parameter $\alpha$ is small enough (here we choose $\alpha=0.02$), the integration range in Eq.~(\ref{phit}) can be extended from $-\infty$ to $\infty$, and the resulting probability density is~\cite{Kim2006,Staelens2021}
\begin{equation}
P(j,t)\approx\frac{1}{\sqrt{\pi}}\frac{1}{\sqrt{\alpha^{-2}+[\alpha\omega_{k_c}''(t+t_c)]^2}}
\exp\left\{-\frac{(j-v_c t)^2}{\alpha^{-2}+[\alpha\omega_{k_c}''(t+t_c)]^2}\right\},
\end{equation}
where $\omega_{k_c}''$ denotes the second derivative of the dispersion relation evaluated at $k_c$. The wave packet maintains the Gaussian shape with the width increasing monotonically over time, thus it spreads and the height diminishes as it propagates towards the giant atom, as shown by the $Jt=-60$ curve in Figs.~\ref{fig5} (a)-(e).

When the wave packet interacts with a giant atom exhibiting loss in Fig.~\ref{fig5} (a), part of it is absorbed. The remaining portion exhibits distinct rapid oscillations arising from interference between the reflected and incident wave packets, and eventually splits into well-separated reflected and transmitted components.
For the case with neither loss nor gain, although one bound state exists in the system, the photon energy lies in the continuum and does not resonate with the discrete bound state. Therefore, the wave packet simply reflects and transmits after interacting with the giant atom, with no localized state formation, as illustrated in Fig.~\ref{fig5} (b).

Completely different wave-packet propagation scenarios appear in the gain regime. In Fig.~\ref{fig5} (c), the wave packet, amplified and modulated by the giant atom with gain $\ga/J=0.167$, continues to reflect and transmit at time $Jt = 160$. Notably, after a long time, a wave packet with exponential tails on both sides appears in the central region, manifesting as a triangular shape on the semilogarithmic plot. This corresponds precisely to the time-growing bound state. Based on Eq.~(\ref{psi}), we have $P(t)\sim e^{2{\rm Im}(E_g) t}$, where $E_g$ denotes the eigenenergy of the time-growing bound state. This demonstrates temporal exponential growth with rate ${\rm Im}\,E_g/J\approx0.016$ shown in the center region. Similarly, it follows from the spatial decay described by $P(j)\sim e^{2{\rm Im}(k_g) j}$ for $j<0$ and $P(j)\sim e^{-2{\rm Im}(k_g) j}$ for $j>N$ that the slopes on both sides of the triangular peak are in excellent agreement with twice the imaginary part of the wave number, i.e., $2{\rm Im}\,k_g\approx0.779$.

A semilogarithmic plot of the probability densities $P(j,t)$ for the critical gain case $\gamma/J\approx0.215$ is displayed in Fig.~\ref{fig5} (d). We find that at early times post-scattering, such as $J t = 160$, the profile develops a broad top-hat-like spatial distribution, a phenomenon known to occur in other systems~\cite{Spectral,Wang2016}. This behavior indicates the emergence of a spectral-singularity state, as we have analyzed in Sec.~\ref{sec4}.
Figure~\ref{Trajectories} (c) shows that the mean wave number of the incoming wave packet exactly matches that of the spectral-singularity state, i.e., $k_c=k_s$. Thus, with $|C_s|^2\gg |C_g|^2$ in Eq.~(\ref{Cn}), the spectral-singularity state dominates the early-time dynamics over the time-growing state. Given that the energy of the spectral-singularity state lies on the real axis, $E_s/J\approx-0.496$, and the corresponding real wave number is $k_s=1.32$, we get $P(j,t)=|\Psi_s(j,t)|\sim \text{constant}$ from Eq.~(\ref{psi}), indicating the bidirectional outgoing plane waves.
However, the persistence of the plateau is limited in time. Since the time-growing state has a positive imaginary part, ${\rm Im}\,E_g/J\approx0.021$, it undergoes sustained amplification, leading to its eventual dominance at later times. In Fig.~\ref{fig5} (d), we observe that the early plateau is replaced by a sharp localized triangular peak with slope $2{\rm Im}\,k_g\approx0.776$ at much later times $Jt = 1940$ and $2240$.

A waveform characterized by a steep central peak with gently sloping wings appears for the case $\gamma/J=0.23$ in Fig.~\ref{fig5} (e). At this gain rate, we can find two time-growing bound states in Figs.~\ref{Trajectories} (c) and (d). Twice the imaginary parts of the wave numbers of these two time-growing bound states, namely $2\operatorname{Im} k_{g_1}\approx0.008$ and $2\operatorname{Im} k_{g_2}\approx0.797$, corresponds precisely to the slopes of the gentle and steep regions, respectively.
All the findings in the gain regime constitute conclusive evidence that the excitation of a time-growing bound state is the direct cause of the system's anomalous long-time behavior.

\section{Conclusions and outlook}\label{con}

In this work, we have investigated the single-photon scattering in a one-dimensional coupled-resonator waveguide, mediated by a giant atom with loss or gain. This system can be modeled by a one-dimensional tight-binding chain, with the giant atom coupled to two lattice sites. To incorporate the effects of the loss and gain, we extend the conventional two-level model for this giant atom by making its on-site energy complex-valued, with its imaginary part $\gamma$ signifying absorption (for $\gamma<0$) or amplification (for $\gamma>0$). The linear gain term $i\gamma$ is a standard effective description,  typically valid below the threshold of the corresponding physical system. Near or above a spectral singularity, gain saturation~\cite{Chong2011}, noise~\cite{Caves1982}, pump dynamics~\cite{Teklu2018}, and quantum-limited amplification~\cite{Clerk2010} may become relevant. These effects require a nonlinear or stochastic treatment, which is beyond the scope of the present analysis. The resulting non-Hermitian Hamiltonian is treated by a correspondingly generalized projection operator formalism, which we then apply to the single-photon scattering problem.

With the scattering theory established on the analyzed scattering states, we derive the expression for the scattering matrix and then present the resulting reflection and transmission coefficients. We find that, unlike the case of small atoms, destructive interference between the two coupling points of the giant atom can decouple the waveguide from the giant atom under certain conditions, resulting in perfect transmission. Moreover, a giant atom with loss absorbs the incident wave, whereas a giant atom with gain not only amplifies it but can also induce divergence at specific energies, which corresponds to persistent wave emission. Solving the eigen equation of the effective Hamiltonian reveals that the divergence corresponds to a spectral-singularity state. Furthermore, numerical analysis of the energy spectrum indicates the simultaneous emergence of at least one time-growing state that exhibits temporal growth. Although the initial overlap of an incident wave packet with the time-growing bound state may be minimal, this component nevertheless experiences exponential amplification, whereby it comes to dominate the long-time evolution. We performed numerical simulations of Gaussian wave-packet propagation to verify the theoretical predictions. In particular, the appearance of the probability density plateau and the subsequent formation of a peak with exponential tails on both sides at long times are attributed to the characteristics of the spectral-singularity states and time-growing states, respectively.
Crucially, our numerical results show that for $\gamma>0$, at least one time-growing bound state exists for all tested parameter combinations. This implies that reflection and transmission coefficients obtained from conventional time-independent scattering methods become unphysical in this regime, so a time-dependent approach is required to properly analyze wave-packet evolution.

Our non-Hermitian giant-atom model can be implemented in two alternative platforms. In superconducting transmon qubits coupled to microwave waveguides, giant atoms are realized by coupling a single transmon to the waveguide at multiple spatially separated points via capacitors, producing tunable atom-waveguide coupling~\cite{Kannan2019}.
The coupling points are separated on-chip by a distance on the order of the wavelength, making the even-odd symmetry of the scattering spectra directly testable by varying the coupling points.
The gain can be introduced by a microwave parametric pump that activates a four-wave mixing process, reducing the transmon-cavity dynamics to an effective gain rate~\cite{Yang2026}, which produces a positive imaginary part in the effective on-site energy. The resulting reflection and transmission spectra can be measured via dispersive readout of the transmon state.
For ultracold atoms in optical lattices, giant atoms can be engineered by coupling a single impurity atom to multiple distant sites of a structured bath, with the waveguide realized using an optical lattice with a tunable band structure~\cite{Tudela2019,Krinner2018,Stewart2020}. The relative gain and loss can be extracted by subtracting the effects of global atom loss following the protocol in Ref.~\cite{Xu2022,Tsuno2025}.

\section*{Acknowledgments}

We acknowledge grant support from the National Natural Science Foundation of China (Grants No. 12475024) and the Shandong Provincial Natural Science Foundation, China (Grants No. ZR2020QA079 and No. ZR2021MA081).

\section*{Data availability statement}
The data that support the findings of this study are openly available~\cite{zhang2026datas}.

\appendix

\section{Calculations of $\langle \tilde{a}|H_{\text{eff}}(E)|a\rangle$ and $\mathcal{R}(k)$}\label{a:cal}

We first present the identity for the following integral:
\begin{equation}
\frac{1}{2\pi}\int_{-\pi}^\pi \frac{e^{iq\ell}}{E-\om_q} dq.
\end{equation}
Substituting the dispersion relation for the waveguide $\om_q=\om_c-2J\cos q$ into above expression, we obtain
\begin{equation}
\frac{1}{2\pi}\int_{-\pi}^\pi \frac{e^{iq\ell}}{E-\om_c+2J\cos q} dq.
\end{equation}
We transform the integral depending on the absolute value $|\ell|$ into an integral over the complex variable $z=e^{i q}$ around the unit circle in the counterclockwise direction,
\begin{equation}
\frac{1}{2\pi}\int_{-\pi}^\pi \frac{e^{iq\ell}}{E-\om_c+2J\cos q} dq
=-\frac{i}{2\pi J}\oint_{|z| =1} \frac{z^{|\ell|}}{(z-e^{i p}) (z-e^{-i p})}d z,
\end{equation}
by introducing the complex momentum $p$ satisfying the same dispersion relation,
\begin{equation}
E=\om_c-2J\cos p.
\end{equation}
For ${\rm Im}\ p>0$ (${\rm Im}\ p<0$), the pole at $e^{i p}$ ($e^{-i p}$) is inside the contour, and the corresponding residue is
\begin{equation}
{\rm Res}\left[-\frac{i}{2\pi J}\oint_{\vert z\vert =1} \frac{z^{\vert \ell\vert }}{ (z-e^{i p}) (z-e^{-i p})}d z,e^{\pm i p}\right]
=\mp\frac{e^{\pm i p|\ell| }}{4\pi J \sin p}.
\end{equation}
The integral is just $2\pi i$ times the residue of the pole by residue theorem, namely
\begin{equation}\label{int}
\frac{1}{2\pi}\int_{-\pi}^\pi \frac{e^{iq\ell}}{E-\om_q} dq
=\mp\frac{i e^{\pm i p|\ell| }}{2 J \sin p}.
\end{equation}

It should be noted that when ${\rm Im}\,p=0$, $E$ lies within the continuum, i.e., $E$ is real and satisfies $\om_c-2J \le E \le \om_c+2J$. In this case, we write $E=\om_k=\om_c-2J\cos k$, and the integral is not well defined in the usual sense. Thus, $\om_k$ is defined via its analytic continuation $\om_k\to\om_k\pm i\epsilon$ (with $\epsilon > 0$) to shift the pole off the real axis into the complex plane. For the right-going wave considered in our work ($0\le k\le \pi$), the corresponding complex momentum thus has a corresponding infinitesimal imaginary part, i.e., $k\to k\pm i\epsilon$, and we obtain
\begin{equation}\label{int0}
\frac{1}{2\pi}\int_{-\pi}^\pi \frac{e^{iq \ell}}{\om_k-\om_q\pm i\epsilon} dq
=\mp\frac{i e^{\pm i k|\ell| }}{2 J \sin k},
\end{equation}
which has exactly the same form as the result in Eq.~(\ref{int}).

Using the Eqs.~(\ref{int}) and (\ref{int0}) , we sequently obtain the effective matrix element:
\begin{equation}
\begin{aligned}
\langle \tilde{a}|H_{\text{eff}}(E)|a\rangle
={}&\langle \tilde{a}|H_{QQ} + H_{QP} \frac{1}{E - H_{PP}} H_{PQ}|a\rangle\\
={}&\langle \tilde{a}|H_{QQ}|a\rangle + \int_{-\pi}^\pi \frac{\langle \tilde{a}|H_{QP}|q\ra\la q|H_{PQ}|a\rangle}{E - \om_q} dq\\
={}&\omega_a +i\gamma + \frac{g^2}{2\pi}\int_{-\pi}^\pi \frac{|1+e^{i q N}|^2}{E - \om_q}dq \\
={}&\begin{dcases*}
\omega_a+i\gamma-\frac{i g^2 (1+e^{i p N })}{J \sin p}, & ${\rm Im}\ p> 0$,\\
\omega_a+i\gamma+\frac{i g^2 (1+e^{-i p N })}{J \sin p}, & ${\rm Im}\ p< 0$.
\end{dcases*}
\end{aligned}
\end{equation}
Replacing $E$ by the discrete eigenvalue $E_n=\omega_c-2J\cos k_n$, with corresponding complex wave numbers $k_n$, yields the result given by Eq.~(\ref{eigenvalue}).

For the case where the energy lies within the continuum, the evaluation of $\mathcal{R}(k)$ in Eq.~(\ref{k}) is given by
\begin{equation}\label{mRk}
\begin{aligned}
\mathcal{R}(k)={}&\langle \tilde{a}|\frac{1}{\omega_k-H_{\text{eff}}(\omega_k+i\epsilon)}|a\rangle
=\frac{1}{\langle \tilde{a}|\omega_k-H_{\text{eff}}(\omega_k+i\epsilon)|a\rangle}\\
={}&-\frac{J\sin k}{J\sin k(\om_a-\om_k+i\gamma)-i g^2(1+e^{ikN})}.
\end{aligned}
\end{equation}
Replacing the effective Hamiltonian with its Hermitian conjugates but keeping the analytic continuation direction $i\epsilon$ unchanged, we get
\begin{equation}\label{mtRk}
\tilde{\mathcal{R}}(k)
=\langle \tilde{a}|\frac{1}{\omega_k-H_{\text{eff}}^\dag(\omega_k+i\epsilon)}|a\rangle
=-\frac{J\sin k}{J\sin k(\om_a-\om_k-i\gamma)-i g^2(1+e^{ikN})}.
\end{equation}
These lead to Eqs.~(\ref{Rk}) and (\ref{tRk}) when $\om_a=\om_c$.

\section{The stationary state approach}\label{a:SS}

In addition to the projection operators formalism presented in the main text, the transmission and reflection coefficients can be derived directly using the stationary state solution method. Within the single-excitation subspace, the eigenstate of the system can be expressed as
\begin{equation}
|\Psi\rangle = \Psi(a) |a\rangle +\sum_j \Psi(j)|j\rangle ,
\end{equation}
where $\Psi(a)$ is the atomic excitation amplitude, and $\Psi(j)$ is the probability amplitude at waveguide site $j$. Substituting this ansatz into the stationary Schr{\"o}dinger equation and applying the boundary conditions imposed by the giant atom coupling at sites $j=0$ and $j=N$, we obtain the following set of equations for the amplitudes:
\begin{equation}\label{se}
\begin{dcases}
(\omega_k - \omega_c)\Psi(0) + J[\Psi(-1)+\Psi(1)] - g \Psi(a) = 0, \\
(\omega_k - \omega_c)\Psi(N) + J[\Psi(N-1)+\Psi(N+1)] - g \Psi(a) = 0, \\
(\omega_k - \omega_a-i\gamma)\Psi(a) - g[\Psi(0) + \Psi(N)] = 0,
\end{dcases}
\end{equation}
where $\omega_k = \omega_c - 2J\cos k$ is the waveguide dispersion.
The solutions to Eqs.~(\ref{se}) are plane waves, and the probability amplitude for a photon incident from the left can be considered in the form of
\begin{equation}\label{Psi}
\Psi(j) =
\begin{dcases}
V e^{ikj} + W e^{-ikj}, & j < 0, \\
X e^{ikj} + Y e^{-ikj}, & 0 \le j < N, \\
Z e^{ikj}, & j \ge N.
\end{dcases}
\end{equation}
with $0\le k\le \pi$.
Together with the continuous condition at $j=0$ and $j=N$, which are $V+W=X+Y$ and $X e^{ikN}+Y e^{-ikN}=Z e^{ikN}$, we can solve Eq.~(\ref{se}) and obtain the scattering amplitudes,
\begin{align}
r ={}&\frac{W}{V}=-\frac{g^2(1+e^{ikN})^2}{2g^2(1+e^{ikN}) - 2iJ\sin k(\omega_k -\omega_a- i\gamma)}, \\
t ={}&\frac{Z}{V}=1 + \frac{g^2(1+e^{ikN})(1+e^{-ikN})}{2iJ\sin k(\omega_k -\omega_a- i\gamma) - 2g^2(1+e^{ikN})}.
\end{align}
One can verify that these amplitudes coincide with Eqs.~(\ref{r}) and (\ref{t}) in the main text for $\omega_a=\omega_c$.

\section{Biorthogonality proof}\label{a:b}
\subsection{Biorthogonality and normalization of bound states}
Taking the bra of the Eq.~(\ref{eqmug}) gives
\begin{equation}
\langle\tilde{\Psi}_n|
=
\mathcal{N}_n\left[
\langle \tilde{a}|
+
\frac{g}{\sqrt{2\pi}}
\int_{-\pi}^{\pi}
\frac{1+e^{ikN}}{E_n-\omega_k}\langle k|\,dk
\right].
\end{equation}
Using \(\langle \tilde{a}|a\rangle=1\), \(\langle \tilde{a}|k\rangle=0\), \(\langle k|a\rangle=0\), and \(\langle k|k'\rangle=\delta(k-k')\), we obtain
\begin{equation}
\langle\tilde{\Psi}_m|\Psi_n\rangle
=
\mathcal{N}_m\mathcal{N}_n
\left[
1-
\mathcal I_{mn}
\right],
\end{equation}
where
\begin{equation}\label{Imn}
\mathcal I_{mn}=
-\frac{g^2}{2\pi}
\int_{-\pi}^{\pi}
\frac{|1+e^{i q N}|^2}
{(E_m-\omega_q)(E_n-\omega_q)}\,dq.
\end{equation}

For $E_m\neq E_n$, we use partial fractions to obtain
\begin{equation}
\mathcal I_{mn}=
\frac{g^2}{2\pi(E_m-E_n)}
\int_{-\pi}^{\pi}
\left[
\frac{|1+e^{i q N}|^2}{E_m-\omega_q}
-\frac{|1+e^{i q N}|^2}{E_n-\omega_q}
\right]dq.
\end{equation}
Substituting the bound-state condition Eq.~(\ref{eigenvalue}),
\begin{equation}
\mathcal I_{mn}=
\frac{(E_m-\omega_a-i\gamma)-(E_n-\omega_a-i\gamma)}{E_m-E_n}
=1,
\end{equation}
and $\langle\tilde{\Psi}_m|\Psi_n\rangle=0$.

For $E_m=E_n$, the normalization condition $\langle\tilde{\Psi}_n|\Psi_n\rangle=1$ gives
\begin{equation}
\mathcal{N}_n
=
\left(
1+
\frac{g^2}{2\pi}
\int_{-\pi}^{\pi}
\frac{|1+e^{i q N}|^2}
{(E_n-\omega_q)^2}\,dq
\right)^{-\frac{1}{2}}.
\end{equation}
Hence,
\begin{equation}\label{D6}
\langle\tilde{\Psi}_m|\Psi_n\rangle=\delta_{mn}.
\end{equation}

\subsection{Biorthogonality and normalization of scattering states}

Taking the Hermitian conjugate of Eq.~(\ref{k+}) gives
\begin{equation}
\langle \tilde\Psi(k)|
=
\frac{g(1+e^{-ikN})}{\sqrt{2\pi}}\,\tilde{\mathcal R}^*(k)\,\langle \tilde a|
+\langle k|
+\frac{g^2(1+e^{-ikN})}{2\pi}\,\tilde{\mathcal R}^*(k)
\int_{-\pi}^{\pi} dq\,
\frac{1+e^{iqN}}{\omega_{k}-\omega_q-i\epsilon}\,\langle q| .
\end{equation}
The inner product of the biorthogonal pairing is
\begin{equation}\label{D9}
\begin{aligned}
\langle \tilde\Psi(k')|\Psi(k)\rangle
={}&
\delta(k-k')
+\frac{g^2(1+e^{-ik'N})(1+e^{ikN})}{2\pi}
\tilde{\mathcal R}^*(k')\mathcal R(k)\\
&\times\left\{1+\left[\frac{1}{\tilde{\mathcal R}^*(k')}-\frac{1}{\mathcal R(k)}\right]\frac{1}{\omega_k-\omega_{k'}+i\epsilon}-\mathcal I_{k'k}\right\},
\end{aligned}
\end{equation}
where
\begin{equation}
\mathcal I_{k'k}
=-\frac{g^2}{2\pi}
\int_{-\pi}^{\pi}
\frac{|1+e^{i q N}|^2}
{(\om_{k'}-\omega_q-i\epsilon')(\om_k-\omega_k+i\epsilon)}\,dq.
\end{equation}
Using the Sokhotski-Plemelj formula,
\begin{equation}
\frac{1}{E_n-\omega_q\pm i\epsilon}
=
\mathcal{P}\frac{1}{E_n-\omega_q}
\mp i\pi\delta(E_n-\omega_q),
\end{equation}
and the principal-value product identity~\cite{Barnett2002},
\begin{equation}
\frac{\mathcal{P}}{E_m-\omega_q}\,
\frac{\mathcal{P}}{E_n-\omega_q}
=
\frac{\mathcal{P}}{E_m-E_n}
\left(
\frac{\mathcal{P}}{E_n-\omega_q}
-
\frac{\mathcal{P}}{E_m-\omega_q}
\right)
+
\pi^2\delta(E_m-\omega_q)\delta(E_n-\omega_q),
\end{equation}
after lengthy calculations, we obtain
\begin{equation}
\begin{aligned}
\mathcal I_{k'k}
={}&\mathcal{P}\frac{1}{\om_k'-\om_k}
\left[\frac{g^2}{2\pi}\int_{-\pi}^\pi\frac{|1+e^{iq N}|^2}{\om_k'-\om_q+i\epsilon}dq
-\frac{g^2}{2\pi}\int_{-\pi}^\pi\frac{|1+e^{iq N}|^2}{\om_k-\om_q+i\epsilon}dq\right]\\
&-\frac{\pi g^2|1+e^{i k N}|^2}{J\sin k}\,
\delta(\om_k'-\om_k)\\
={}&\mathcal{P}\frac{1}{\om_{k'}-\om_k}
\left[\frac{i g^2(1+e^{-ik' N})}{J\sin k'}+\frac{i g^2(1+e^{ik N})}{J\sin k}
\right]
-\frac{\pi g^2|1+e^{i k N}|^2}{J\sin k}\,
\delta(\om_{k'}-\om_k).
\end{aligned}
\end{equation}
The second term within the braces on the right-hand side of Eq.~(\ref{D9}) yields
\begin{equation}
\begin{aligned}
&\left[\frac{1}{\tilde{\mathcal R}^*(k')}-\frac{1}{\mathcal R(k)}\right]\frac{1}{\omega_k-\omega_{k'}+i\epsilon}\\
={}&\left[\frac{1}{\tilde{\mathcal R}^*(k')}-\frac{1}{\mathcal R(k)}\right]\mathcal{P}\frac{1}{\omega_k-\omega_{k'}}
-i\pi\left[\frac{1}{\tilde{\mathcal R}^*(k')}-\frac{1}{\mathcal R(k)}\right]\delta(\omega_k-\omega_{k'})\\
={}&-1+\mathcal{P}\frac{1}{\omega_{k'}-\omega_k}\left[\frac{i g^2(1+e^{-ik' N})}{J\sin k'}+\frac{i g^2(1+e^{ik N})}{J\sin k}\right]\\
&-\frac{\pi g^2|1+e^{i k N}|^2}{J\sin k}\delta(\omega_k-\omega_{k'}),
\end{aligned}
\end{equation}
cancelling the other two terms in the brace in Eq.~(\ref{D9}). We are left with
\begin{equation}
\langle \tilde\Psi(k')|\Psi(k)\rangle=\delta(k-k').
\end{equation}

\subsection{Biorthogonality between the scattering state and the bound state}

Here we prove the orthogonality between the scattering state \(|\Psi(k)\rangle\) and the bound state \(|\Psi_n\rangle\) in the biorthogonal sense. Substituting the expansions into the inner product and using the pairing rules, we have
\begin{equation}\label{D16}
\begin{aligned}
\langle\tilde\Psi_n|\Psi(k)\rangle
={}&\frac{g\mathcal{N}_n(1+e^{ikN})}{\sqrt{2\pi}}
\left[
\mathcal R(k)
+\frac{1}{E_n-\omega_k-i\epsilon}
-\mathcal R(k)\mathcal I_{nk}\right]\\
={}&\frac{g\mathcal{N}_n(1+e^{ikN})}{\sqrt{2\pi}}\mathcal R(k)
\left[1
+\frac{1}{\mathcal R(k)}\frac{1}{E_n-\omega_k-i\epsilon}
-\mathcal I_{nk}\right],
\end{aligned}
\end{equation}
where
\begin{equation}
\mathcal I_{nk}
=-\frac{g^2}{2\pi}
\int_{-\pi}^{\pi}
\frac{|1+e^{i q N}|^2}
{(E_n-\omega_q)(\om_k-\omega_k+i\epsilon)}\,dq.
\end{equation}
Since the scattering-state energy cannot be equal to the bound-state energy, the third term in the square brackets reduces to
\begin{equation}
\begin{aligned}
\mathcal I_{nk}
={}&\frac{g^2}{2\pi(E_n-\om_k-i\epsilon)}
\int_{-\pi}^{\pi}
\left[
\frac{|1+e^{i q N}|^2}{E_n-\omega_q}
-\frac{|1+e^{i q N}|^2}{\om_k-\omega_q+i\epsilon}
\right]dq\\
={}&\frac{1}{E_n-\om_k-i\epsilon}\left[E_n-\om_a-i\ga +\frac{i g^2 (1+e^{i k N })}{J \sin k}\right]\\
={}&1+\frac{1}{\mathcal R(k)}\frac{1}{E_n-\omega_k-i\epsilon}.
\end{aligned}
\end{equation}
Thus the expression in the square brackets in Eq.~(\ref{D16}) vanishes identically, and consequently,
\begin{equation}
\langle\tilde\Psi_n|\Psi(k)\rangle=0.
\end{equation}

\section{Completeness relation proof}\label{a:i}
The new resolution of identity is
\begin{equation}
\sum_n|\Psi_n\rangle\langle\tilde{\Psi}_n|
+\int_{-\pi}^{\pi}dk\,|\Psi(k)\rangle\langle\tilde{\Psi}(k)|
=I,
\end{equation}
where the summation in the first term on the left-hand side is taken over the discrete states satisfying the bound-state condition Eq.~(\ref{eigenvalue}), while the integral in the second term contains all scattering states whose energies along the branch.
This is precisely the completeness relation used in Eq.~(\ref{phi0}) and it is valid under the assumption that the spectrum is non-degenerate. In what follows, we prove that the states $|\Psi_n\rangle,|\Psi(k)\rangle$ and their dual states $|\tilde{\Psi}_n\rangle,|\tilde{\Psi}(k)\rangle$ satisfy the biorthogonal completeness relation.

Since
$Q+P=I,QP=PQ=0,Q=|a\rangle\langle\tilde{a}|,P=\int_{-\pi}^{\pi}dk\,|k\rangle\langle k|$, the basis \(\{|a\rangle,|k\rangle\}\) is complete.
We verify the matrix elements of
\begin{equation}
\Pi=\sum_n|\Psi_n\rangle\langle\tilde{\Psi}_n|
+\int_{-\pi}^{\pi}dk\,|\Psi(k)\rangle\langle\tilde{\Psi}(k)|
\end{equation}
in this basis, and the completeness therefore suffices to show that
\begin{equation}
\begin{aligned}
&\langle\tilde{a}|\Pi|a\rangle=1,
&&\langle\tilde{a}|\Pi|k\rangle=0,\\
&\langle k|\Pi|a\rangle=0,
&&\langle k'|\Pi|k\rangle=\delta(k-k').
\end{aligned}
\end{equation}

\subsection{$\langle\tilde{a}|\Pi|a\rangle=1$}

Consider the contour integral in the first Riemann sheet,
\begin{equation}
\frac{1}{2\pi i}\oint_{C_\infty}\frac{dE}{E-\Omega(E)},
\end{equation}
with
\begin{equation}
\Omega(E)=\om_a+i\gamma+\frac{g^2}{2\pi}\int_{-\pi}^\pi \frac{|1+e^{i k N}|^2}{E-\om_k}dk.
\end{equation}
The contour $C_\infty$ is taken at infinity in the counterclockwise direction.
When $|E|\to\infty$, $\Omega(E)\to \om_a+i\ga$, which is bounded. Hence
\begin{equation}
\frac{1}{2\pi i}\oint_{C_\infty}\frac{dE}{E-\Omega(E)}=\frac{1}{2\pi i}\oint_{C_\infty}\frac{1}{E}\sum_{j=0}^\infty\left[\frac{\Omega(E)}{E}\right]^j
=1.
\end{equation}
By deforming $C_\infty$ onto the isolated poles $E_n$ and the branch cut $[\omega_c-2J,\omega_c+2J]$, one obtains
\begin{equation}
\frac{1}{2\pi i}\oint_{C_\infty}\frac{dE}{E-\Omega(E)}=\sum_{n}{\rm Res}\left[\frac{1}{E-\Omega(E)},E_n\right]+\frac{1}{2\pi i}\oint_{C_{\rm cut}}\frac{dE}{E-\Omega(E)},
\end{equation}
with the contour $C_{\rm cut}$ running counterclockwise along both sides of the cut.
At the poles $E_n=\Om(E_n)$,
\begin{equation}
{\rm Res}\left[\frac{1}{E-\Omega(E)},E_n\right]
=\frac{1}{1-\Omega'(E_n)}=\mathcal N_n^2,
\end{equation}
with $\mathcal{N}_n$ obtained in Eq.~(\ref{D6}).
The small arcs around the branch points yield vanishing contributions in the limit of vanishing radius and are omitted. The cut contribution becomes the discontinuity across the cut:
\begin{equation}
\begin{aligned}
\frac{1}{2\pi i}\oint_{C_{\rm cut}}\frac{dE}{E-\Omega(E)}
={}&\frac{1}{2\pi i}\int_{\om_c-2J}^{\om_c+2J}\left[\frac{1}{\om_k-\Omega(\om_k-i\epsilon)} -\frac{1}{\om_k-\Omega(\om_k+i\epsilon)}\right]d\omega_k\\
={}&\frac{g^2}{2\pi}\int_{-\pi}^{\pi}\frac{|1+e^{ikN}|^2}{[\om_k-\Omega(\om_k-i\epsilon)] [\om_k-\Omega(\om_k+i\epsilon)]} dk\\
={}&\frac{g^2}{2\pi}\int_{-\pi}^{\pi}
|1+e^{ikN}|^2
\mathcal{R}(k)\tilde{\mathcal{R}}^*(k)dk,
\end{aligned}
\end{equation}
where we use
\begin{equation}
\Omega(\om_k\pm i\epsilon)=\omega_a+i\gamma\mp\frac{i g^2 (1+e^{\pm i k N })}{J \sin k},
\end{equation}
together with the explicit forms of $\mathcal{R}(k)$ and $\tilde{\mathcal{R}}(k)$ in Eqs.~(\ref{mRk}) and (\ref{mtRk}).Using the state definitions,
\begin{equation}
\begin{aligned}
&\langle\tilde{a}|\Psi_n\rangle=\mathcal{N}_n,
&&\langle\tilde{\Psi}_n|a\rangle=\mathcal{N}_n,\\
&\langle\tilde{a}|\Psi(k)\rangle
=
\frac{g(1+e^{ikN})}{\sqrt{2\pi}}\mathcal{R}(k),\qquad
&&\langle\tilde{\Psi}(k)|a\rangle
=
\frac{g(1+e^{-ikN})}{\sqrt{2\pi}}\tilde{\mathcal{R}}^*(k),
\end{aligned}
\end{equation}
we obtain
\begin{equation}
\begin{aligned}
\langle\tilde{a}|\Pi|a\rangle
={}&\sum_n \mathcal{N}_n^2
+\frac{g^2}{2\pi}
\int_{-\pi}^{\pi}
|1+e^{ikN}|^2
\mathcal{R}(k)\tilde{\mathcal{R}}^*(k)dk\\
={}&\sum_{n}{\rm Res}\left[\frac{1}{E-\Omega(E)},E_n\right]+\frac{1}{2\pi i}\oint_{C_{\rm cut}}\frac{dE}{E-\Omega(E)}\\
={}&1.
\end{aligned}
\end{equation}

\subsection{$\langle\tilde{a}|\Pi|k\rangle=0$ and $\langle k|\Pi|a\rangle=0$}

We first consider
\begin{equation}
\langle\tilde{a}|\Pi|k\rangle
=\frac{g(1+e^{ikN})}{\sqrt{2\pi}}\left[
\sum_n\frac{\mathcal{N}_n^2}{E_n-\omega_k-i\epsilon}
+\mathcal{R}(k)
+\frac{g^2}{2\pi}\int_{-\pi}^{\pi}
\frac{(1+e^{iqN})(1+e^{-iqN})
\mathcal{R}(q)\tilde{\mathcal{R}}^*(q)}
{\omega_q-\omega_k-i\epsilon}dq\right].
\end{equation}
We turn to the contour integral in the first Riemann sheet,
\begin{equation}
\frac{1}{2\pi i}\oint_{C_\infty}
\frac{dE}{E-\Omega(E)}\,
\frac{1}{E-\omega_k-i\epsilon}.
\end{equation}
Since the integrand decays as $E^{-2}$ for large $|E|$, the contribution from the circle at infinity vanishes. Deforming the contour onto the isolated poles $E_n$ and $\omega_k+i\epsilon$, and the branch cut, we evaluate each contribution.

At the isolated poles \(E_n=\Omega(E_n)\),
\begin{equation}
{\rm Res}\left[\frac{1}{E_n-\omega_k-i\epsilon}\frac{1}{E-\Omega(E)},E_n\right]
=
\frac{1}{E_n-\omega_k-i\epsilon}\frac{1}{1-\Omega'(E_n)}
=
\frac{\mathcal{N}_n^2}{E_n-\omega_k-i\epsilon}.
\end{equation}
The pole at $E=\omega_k+i\epsilon$, arising from $1/(E-\omega_k-i\epsilon)$, yields the residue
\begin{equation}
{\rm Res}\left[\frac{1}{E_n-\omega_k-i\epsilon}\frac{1}{E-\Omega(E)},\omega_k+i\epsilon\right]
=\frac{1}{\omega_k+i\epsilon-\Omega(\omega_k+i\epsilon)}
=\mathcal{R}(k).
\end{equation}
Along the branch cut we set \(E=\omega_q\pm i\epsilon'\) with an independent infinitesimal \(\epsilon'<\epsilon\). The discontinuity across the cut gives
\begin{equation}
\frac{1}{2\pi i}\oint_{C_{\rm cut}}
\frac{dE}{E-\Omega(E)}
\frac{1}{E-\omega_k-i\epsilon}
=
\frac{g^2}{2\pi}\int_{-\pi}^{\pi}
\frac{(1+e^{iqN})(1+e^{-iqN})
\mathcal{R}(q)\tilde{\mathcal{R}}^*(q)}
{\omega_q-\omega_k-i\epsilon}dq.
\end{equation}
Collecting all contributions, we obtain
\begin{equation}
\sum_n\frac{\mathcal{N}_n^2}{E_n-\omega_k-i\epsilon}
+\mathcal{R}(k)
+\frac{g^2}{2\pi}\int_{-\pi}^{\pi}
\frac{(1+e^{iqN})(1+e^{-iqN})
\mathcal{R}(q)\tilde{\mathcal{R}}^*(q)}
{\omega_q-\omega_k-i\epsilon}dq
=0.
\end{equation}
Hence
\begin{equation}
\langle\tilde{a}|\Pi|k\rangle=0.
\end{equation}
Taking the limits $\epsilon'\to0^+$ and then $\epsilon\to0^+$ completes the proof.

The proof of $\langle k|\Pi|a\rangle=0$ proceeds in exactly the same manner. Inserting the state definitions gives
\begin{equation}
\langle k|\Pi|a\rangle
=
\frac{g(1+e^{-ikN})}{\sqrt{2\pi}}\,
\left[\sum_n\frac{\mathcal{N}_n^2}{E_n-\omega_k+i\epsilon}
+\tilde{\mathcal{R}}^*(k)
+\frac{g^2}{2\pi}\int_{-\pi}^{\pi}
\frac{(1+e^{iqN})(1+e^{-iqN})
\mathcal{R}(q)\tilde{\mathcal{R}}^*(q)}
{\omega_q-\omega_k+i\epsilon}dq\right].
\end{equation}
The same contour integral, now with $1/(E-\omega_k+i\epsilon)$, yields the residues $\mathcal{N}_n^2/(E_n-\omega_k+i\epsilon)$ at $E_n$, $\tilde{\mathcal{R}}^*(k)$ at $E=\omega_k-i\epsilon$, and the corresponding cut contribution. Their sum again vanishes, so
\begin{equation}
\langle k|\Pi|a\rangle=0.
\end{equation}
Thus both matrix elements vanish, as required.

\subsection{$\langle k'|\Pi|k\rangle=\delta(k-k')$}

Substituting the state definitions into $\langle k'|\Pi|k\rangle$, we obtain
\begin{equation}
\begin{aligned}
\langle k'|\Pi|k\rangle
={}&
\delta(k-k')
+\frac{g^2}{2\pi}(1+e^{-ik'N})(1+e^{ikN})
\Bigg[
\sum_n
\frac{\mathcal{N}_n^2}
{(E_n-\omega_{k'}+i\epsilon)(E_n-\omega_k-i\epsilon)}
\\
&\qquad
+\frac{\tilde{\mathcal{R}}^*(k')}{\omega_{k'}-\omega_k-i\epsilon}
+\frac{\mathcal{R}(k)}{\omega_k-\omega_{k'}+i\epsilon}
\\
&\qquad
+\frac{g^2}{2\pi}
\int_{-\pi}^{\pi}
\frac{(1+e^{iqN})(1+e^{-iqN})
\mathcal{R}(q)\tilde{\mathcal{R}}^*(q)}
{(\omega_q-\omega_{k'}+i\epsilon)(\omega_q-\omega_k-i\epsilon)}dq
\Bigg].
\end{aligned}
\end{equation}
It remains to show that the bracketed expression vanishes. To this end, consider the contour integral
\begin{equation}
\frac{1}{2\pi i}\oint_{C_\infty}
\frac{dE}{E-\Omega(E)}
\frac{1}{E-\omega_k-i\epsilon}
\frac{1}{E-\omega_{k'}+i\epsilon}.
\end{equation}
Since the integrand decays as $E^{-3}$ at infinity, the contour at infinity contributes nothing. Deforming the contour onto the isolated poles $E_n$, the shifted poles $E=\omega_k+i\epsilon$ and $E=\omega_{k'}-i\epsilon$, and the branch cut, the sum of all contributions vanishes. The branch-cut integral is evaluated with an independent infinitesimal \(\epsilon'<\epsilon\) to avoid double counting. Consequently, the bracketed expression is zero, and therefore
\begin{equation}
\langle k'|\Pi|k\rangle=\delta(k-k').
\end{equation}
Taking the limits \(\epsilon'\to0^+\) and then \(\epsilon\to0^+\) completes the proof.

\section{Long-time asymptotic via the stationary phase method}\label{a:lt}

We consider the integral
\begin{equation}
I(t)=\int_{-\pi}^{\pi} D(k)\langle j|\Psi(k)\rangle e^{-i\omega_k t}dk,
\end{equation}
where $D(k)$ is defined in Eq.~(\ref{Dk}) and
\begin{equation}\label{F2}
\begin{aligned}
\langle j|\Psi(k)\rangle
={}&\frac{1}{\sqrt{2\pi}}\left[e^{i k j} +\dfrac{g^2(1+e^{ikN})}{2\pi}\mathcal{R}(k)\int_{-\pi}^\pi\dfrac{(1+e^{-iqN})e^{i q j}}{\om_k-\om_q+i\epsilon} dq\right]\\
={}&\frac{1}{\sqrt{2\pi}}\left[e^{i k j} +\dfrac{i g^2(1+e^{ikN})}{2J\sin k}\mathcal{R}(k)\left(e^{i k |j|}+e^{i k |j-N|}\right)\right].
\end{aligned}
\end{equation}
The amplitude is smooth and bounded on the real axis, except possibly at the stationary points of $\omega_k$, which we analyze first.

The stationary points satisfy
\begin{equation}
\omega'_{k_s}=-2J\sin k_s=0,
\end{equation}
hence
\begin{equation}
k_s=0,\;\pm\pi.
\end{equation}
The second derivative is
\begin{equation}
\omega''_{0}=-\omega''_{\pm\pi}=2J>0.
\end{equation}
Thus the stationary points are non-degenerate. We will proceed to verify that $D(k_s)\langle j|\Psi(k_s)\rangle$ is finite.

If \(1+e^{ik_sN}\neq0\), which includes $k_s=0$ and $k_s=\pm\pi$ with $N$ even, then $\mathcal{R}(k)=0$, and $\mathcal{R}(k)/\sin k$ tends to a finite constant.
If $k_s=\pi$ and $N$ is odd, then $1+e^{\pm i\pi N}=0$. Writing $k_s=\pm\pi\mp \eta$, we have
\begin{equation}
1+e^{\pm i\pi N}=1-e^{\mp i\eta N}\approx\pm i \eta N,
\end{equation}
and
\begin{equation}
\sin k_s\approx \mp \eta.
\end{equation}
$\mathcal{R}(k_s)$ in Eq.~(\ref{mRk}) becomes
\begin{equation}
\mathcal{R}(k_s)\approx-\frac{J}{J(\omega_a-\omega_c-2J+i\gamma)+g^2N}.
\end{equation}
Moreover,
\begin{equation}
\dfrac{(1+e^{ik_sN})}{2J\sin k}\mathcal{R}(k_s)
\approx  -\frac{i N J}{J(\omega_a-\omega_c-2J+i\gamma)+g^2N},
\end{equation}
which is also finite. Therefore the $\langle j|\Psi(k_s)\rangle$ is finite at $k_s$ in all cases. The same holds for in $D(k_s)$.

Since $D(k_s)\langle j|\Psi(k_s)\rangle$ is bounded for all stationary points, we may expand $\omega_k$ locally:
\begin{equation}
\omega_k=\omega_{k_s}+\frac{1}{2}\omega''_{k_s}(k-k_s)^2+O[(k-k_s)^3],
\end{equation}
and
\begin{equation}
e^{-i\omega_k t}
=e^{-i\omega_{k_s}t}\,
e^{-i\frac{1}{2}\omega''_{k_s}t\,(k-k_s)^2}
\left\{1+O[(k-k_s)^3t]\right\}.
\end{equation}
The contribution from a small neighborhood of $k_s$ is therefore
\begin{equation}
\begin{aligned}
I_s(t)={}&D(k_s)\langle j|\Psi(k_s)\rangle e^{-i\omega_{k_s}t}
\int_{-\infty}^{\infty}
e^{-i\frac{1}{2}\omega''_{k_s}t\,(k-k_s)^2} d(k-k_s)+O\left(\frac{1}{t}\right)\\
={}&D(k_s)\langle j|\Psi(k_s)\rangle
e^{-i\omega_{k_s}t}
\sqrt{\frac{2\pi}{ |\omega''_{k_s}|t}}\,
e^{-i\frac{\pi}{4}\operatorname{sgn}\left(\omega''_{k_s}\right)}
+O\left(\frac{1}{t}\right).
\end{aligned}
\end{equation}
Thus each stationary point contributes a term of order $t^{-1/2}$.

Away from the stationary points, integration by parts yields a contribution of order $O\left(1/t\right)$, which is negligible compared with the stationary-point contributions.
Hence
\begin{equation}
\lim_{t\to\infty}I(t)=0.
\end{equation}

\section{The Siegert boundary condition}\label{a:sb}

When $V$ equals zero in Eq.~(\ref{Psi}), poles emerge in the scattering amplitudes Eqs.~(\ref{r}) and (\ref{t}). One should apply the Siegert boundary condition to describe these states. The Siegert boundary condition expands the discrete set of $k_n$ to the complex plane, with associated complex energies $E_n =\omega_c-2J\cos k_n$. The time evolution of the corresponding eigenstate is given by
\begin{equation}
\Psi(j,t) = e^{-iE_n t}\times
\begin{dcases}
W e^{-ik_nj}, & j < 0, \\
X e^{ik_nj} + Y e^{-ik_nj}, & 0 \le j < N, \\
Z e^{ik_nj}, & j \ge N.
\end{dcases}
\end{equation}

The states are classified according to the position of $k_n$ in the complex plane. $k_n$ on the positive and negative parts of the lines ${\rm Re}\,k=0$ and ${\rm Re}\,k=\pi$ correspond to conventional bound states and virtual states, respectively. The wave numbers of the resonant states lie in the fourth quadrant, while those of their antiresonant partners lie in the third quadrant. Both energies lie on the second Riemann sheet of the complex energy plane, in contrast to the energies considered in the main text, which all lie on the first Riemann sheet. For the non-unitary cases, $k_n$ in the first quadrant gives ${\rm Im}\,E_n>0$, and the norms of these states grow exponentially with time; $k_n$ in the second quadrant gives ${\rm Im}\,E_n<0$, and the norms of these states decrease with time.

\bibliography{giant1}

\end{document}